\documentclass[12pt]{article}

\usepackage{latexsym}
\usepackage{amssymb,amsfonts,amsmath}
\usepackage{graphicx} 
\usepackage{indentfirst}
\usepackage{bbm}
\usepackage{amssymb}
\usepackage{verbatim}
\usepackage{amsmath, amsthm,amssymb}
\usepackage{mathrsfs}
\usepackage{hyperref}
\usepackage{amsfonts}
\usepackage{dsfont}
\usepackage{cite}
\usepackage{tikz}
\usepackage{enumitem}

\parskip=\medskipamount

\newcommand {\cA}{{\cal A}}
\newcommand {\cB}{{\cal B}}

\newcommand {\cD}{{\cal D}}

\newcommand {\cH}{{\cal H}}
\newcommand {\cI}{{\cal I}}
\newcommand {\cJ}{{\cal J}}

\newcommand {\cL}{{\cal L}}

\newcommand {\cN}{{\cal N}}
\newcommand {\cO}{{\cal O}}

\newcommand {\cQ}{{\cal Q}}

\newcommand{\bX}{{\bf X}}

\newcommand{\bZ}{{\bf Z}}
\def\a{\alpha}

\def\b{\beta}

\def\d{\delta}
\def\e{\epsilon}

\def\g{\gamma}

\def\l{\lambda}
\def\m{\mu}

\def\o{\omega}

\def\q{\theta}
\def\r{\rho}
\def\s{\sigma}

\def\x{\xi}

\def\D{\Delta}
\def\F{\Phi}
\def\J{\Psi}
\def\L{\Lambda}

\def\P{\Pi}
\def\Q{\Theta}

\def\ri{{\rm i}}

\def\ra{{\rm a}}

\newcommand{\ad}{{\dot{\alpha}}}                           
\newcommand{\bd}{{\dot{\beta}}}                            
\newcommand{\ve}{\varepsilon}                            

\newcommand{\pa}{\partial}                           
\newcommand{\hf}{\frac12}

\newcommand{\be}{\begin{equation}}
\newcommand{\ee}{\end{equation}}
\newcommand{\bea}{\begin{eqnarray}}
\newcommand{\eea}{\end{eqnarray}}
\newcommand{\non}{\nonumber}
\newcommand{\1}{{\underline{1}}}
\newcommand{\2}{{\underline{2}}}

\newcommand{\bm}[1]{\mbox{\boldmath$#1$}}

\def\double #1{#1{\hbox{\kern-2pt $#1$}}}

\newcommand{\gd}{{\dot\g}}

\newif\ifdtup

\def\la{{\langle}}
\def\ra{{\rangle}}
\def\fn3{{\bm {X}_{3}, \bm {\Q}_{3}, \bar {\bm \Q}_{3}}}
\def\fxq{{\bm {X}, \bm {\Q}, \bar {\bm \Q}}}
\def\corr1{{\la \bar S_{\ad}(z_1) S_{\b}(z_2) J_{\g \gd}(z_3) \ra}}

\newcommand{\bsubeq}{\begin{subequations}}
\newcommand{\esubeq}{\end{subequations}}

\numberwithin{equation}{section}

\begin{document}

\begin{titlepage}
\begin{flushright}
March, 2021 \\
\end{flushright}
\vspace{5mm}

\begin{center}
{\Large \bf 
Correlation functions of spinor current multiplets in ${\cN}=1$ superconformal theory}
\\ 
\end{center}

\begin{center}

{\bf
Evgeny I. Buchbinder, Jessica Hutomo
and Sergei M. Kuzenko} \\
\vspace{5mm}

\footnotesize{
{\it Department of Physics M013, The University of Western Australia\\
35 Stirling Highway, Crawley W.A. 6009, Australia}}  
~\\

\vspace{2mm}
~\\
\texttt{Email: evgeny.buchbinder@uwa.edu.au,
jessica.hutomo@uwa.edu.au, sergei.kuzenko@uwa.edu.au}
\vspace{2mm}

\end{center}

\begin{abstract}
\baselineskip=14pt

We consider ${\cal N}=1$ superconformal field theories in four dimensions possessing 
an additional conserved spinor current multiplet $S_{\alpha}$ and study three-point functions involving such an operator. A conserved spinor current multiplet naturally exists in superconformal theories with ${\cal N}=2$ supersymmetry and contains the current of the second supersymmetry.  However, we do not assume ${\cal N}=2$ supersymmetry. We show that the three-point function of two spinor current multiplets and the ${\cal N}=1$ supercurrent depends on three independent tensor structures and, in general, is not contained in the three-point function of the ${\cal N}=2$ supercurrent. It then follows, based on symmetry considerations only, that the existence of one more Grassmann odd current multiplet in ${\cal N}=1$ superconformal field theory does not necessarily imply  ${\cal N}=2$ superconformal symmetry.

\end{abstract}
\vspace{5mm}

\vfill
\end{titlepage}

\newpage
\renewcommand{\thefootnote}{\arabic{footnote}}
\setcounter{footnote}{0}

\tableofcontents{}
\vspace{1cm}
\bigskip\hrule

\allowdisplaybreaks


\section{Introduction}


The study of the general structure of correlation functions of primary fields
in conformal field theory has a long history. 
Early theoretical ideas and results can be traced back to the 
1970s~\cite{Polyakov:1970xd, Schreier:1971um, Migdal:1971xh, Migdal:1972tk, Ferrara:1972cq, Ferrara:1973yt, Koller:1974ut, Mack:1976pa, Stanev:1988ft}
(see also a review~\cite{Fradkin:1978pp} and references therein). A systematic approach to determining two- and three-point functions
of conserved currents, such as the energy-momentum tensor and vector current, was undertaken by Osborn and Petkou in~\cite{OP} and
by Erdmenger and Osborn in~\cite{EO}. Such correlation functions 
are severely constrained by conformal symmetry and the conservation properties and are fixed up to finitely many coefficients. 
In certain cases these coefficients are related to Weyl or chiral anomalies
of a conformal field theory coupled to an external curved background or an external gauge field.  

Since the late 1990s,  the approach of~\cite{OP, EO} was generalised to (super)conformal field theories in 
diverse dimensions~\cite{Park1, OsbornN1, Park, Park6, Park3, KT, Giombi:2011rz, Nizami:2013tpa, Buchbinder:2015qsa, Buchbinder:2015wia, 
Kuzenko:2016cmf, Buchbinder:2021gwu}. 
In supersymmetric case the energy-momentum tensor and vector currents must be supplemented with 
new fermionic conserved currents. The energy-momentum tensor is then replaced with the supercurrent and 
vector currents are replaced with flavour current multiplets. 
Thus, the supercurrent and flavour current multiplets contain fundamental information about the symmetries of a 
given supersymmetric field theory.
The concept of supercurrent was  introduced in~\cite{Ferrara:1974pz}, while flavour current multiplets were discussed for the first time in \cite{FWZ}. 
For discussions of 
the most general four-dimensional $\cN=1$ non-conformal supercurrents, see~\cite{MSW, KS2, K-var, K-var1}.
Recently, general properties of superconformal theories in diverse dimensions were studied 
in~\cite{Cordova:2015fha, Cordova:2016xhm, Cordova:2016emh, Cordova:2019wns, Cordova:2020tij}. 

Most analysis of correlation functions in (super)conformal field theories has been performed in coordinate space. 
A formalism in momentum space was recently developed in~\cite{Coriano:2012wp, Coriano:2013jba, Bzowski:2013sza, 
Bzowski:2015pba, Bzowski:2015yxv, Bzowski:2017poo, Bzowski:2018fql}.

The aim of this paper is to study four-dimensional $\cN=1$ superconformal field theory possessing a conserved {\it spinor current multiplet}. 
Such a current naturally exists in $\cN=2$ superconformal field theory and is a part of the 
$\cN=2$ supercurrent $\cJ$~\cite{Sohnius:1978pk, HST}. In general, $\cJ$ contains several conserved $\cN=1$ currents. To describe them 
let us split the $\cN=2$ covariant derivatives $(D_{\a}^i, \bar D^{\dot \a}_i)$ into $(D_{\a}^{\1} =D_{\a}, \bar D^{\dot \a}_{\1}= \bar D^{\dot \a})$ and 
$(D_{\a}^{\2}, \bar D^{\dot \a}_{\2})$  and define the 
$\cN=1$ bar-projection $U|= U(x, \theta^{\a}_i, \bar \theta_{\dot \a}^j)|_{\theta_{\2}= \bar \theta^{\2}\,=0}$. Then $\cJ$ is composed 
out of the three independent $\cN=1$ currents~\cite{KT}:
\be 
J= \cJ|\,, \qquad J_{\a}= D_{\a}^{\2} \cJ| \,, \qquad 
J_{\a  \ad} =\frac{1}{2}[ D_{\a}^{\2}, \bar D_{\dot \a \2}] \cJ| -\frac{1}{6}[ D^{\1}_{\a}, \bar D_{\dot \a \1}]\cJ|\,. 
\label{e0}
\ee
Here $J_{\a \dot \a}$ is the $\cN=1$  supercurrent containing 
the energy-momentum tensor, the current of the first supersymmetry and the $\cN=1$ $R$-symmetry current ~\cite{Ferrara:1974pz}. 
The spinor current multiplet $J_{\a}$ contains the current of the second supersymmetry and two of the three $\rm SU(2)$ $R$-symmetry currents.
Finally, $J$ contains the current corresponding 
to the combined $\cN=2$ U(1) $R$-transformations 
and SU(2) $z$-rotations which leave $\q_{\underline{1}}$ 
and ${\bar \q}^{\underline{1}}$ invariant.
In this paper, however, we do not assume $\cN=2$ supersymmetry. 
We study just $\cN=1$ superconformal field theory which also has a Grassmann odd spinor current multiplet $S_{\a}$ and ask the question whether the
existence of such a current implies $\cN=2$ superconformal symmetry. For this purpose we study three-point functions involving $S_{\a}$
and find their most general form consistent with $\cN=1$ superconformal symmetry. If existence of $S_{\a}$ implies $\cN=2$ superconformal symmetry,
such three-point functions should be all obtained from the three-point function  of the $\cN=2$ supercurrent $\cJ$ by superspace reduction.

A surprising result is that the answer to the above question is negative. More precisely, we find that the three point function
\be
\la \bar S_{\ad}(z_1) S_{\b}(z_2) J_{\g \gd}(z_3) \ra
\label{e00}
\ee
is, in general, inconsistent with $\cN=2$ superconformal symmetry because it has more independent tensor structures. 
That is, despite the fact that $S_{\a}$ in components contains a conserved spin-$\frac{3}{2}$ current, its existence does not
imply $\cN=2$ superconformal symmetry. We would like to stress that our results are based on the symmetry considerations only
and we do not know how to realise them in local field theory. 
We were not able to find explicit field theoretic examples of  $\cN=1$ superconformal theories
possessing a spinor current multiplet, yet without being $\cN=2$ supersymmetric. 
On the other hand, we demonstrate that the three-point function 
\be
\la \bar{S}_{\ad} (z_1) S_{\b} (z_2) L(z_3) \ra~, \ee
where $L$ is a flavour current superfield in an ${\cN}=1$ superconformal theory, is fixed up to two real parameters. This result is in agreement with ${\cN}=2$ superconformal symmetry, in the sense that the three-point function $\la \bar{J}_{\ad}(z_1) J_{\b}(z_2) J(z_3) \ra$ also possesses two linearly independent structures. 
 
One natural question which arises in this context is whether our results are intrinsic for $\cN=1$ superconformal symmetry 
or also hold in non-supersymmetric case. That is, let us consider a non-supersymmetric conformal theory 
which also possesses a conserved Grassmann odd spin-$\frac{3}{2}$ current $Q_{m \a}$. Based on symmetries only, does the existence 
of $Q_{m \a}$ imply $\cN=1$ supersymmetry? If the answer is positive the three-point function
\be
\la \bar Q_{m \dot \a}(x_1) Q_{n \b}(x_2) T_{p q}(x_3) \ra\,, 
\label{e000}
\ee
where $T_{pq}$ is the energy-momentum tensor, must be contained in the three-point function of the $\cN=1$ supercurrent. 
This question is left for future research.\footnote{To best of our knowledge 
this correlation function has not been previously studied.} One more natural question is whether our results are intrinsic to four-dimensional 
(super)conformal symmetry or hold in say three and six dimensional (super)conformal theories. This question is also left for future research.

The paper is organised as follows. In section \ref{Section2} we review the general construction of three-point functions in superconformal theories. 
In our review we follow~\cite{OsbornN1, Park1, KT, Park} where additional details can be found. In section \ref{Section3} we discuss three-point 
functions of spinor current multiplets and the supercurrent in $\cN=1$ superconformal theories. Our main result is that the three point function~\eqref{e00}
is fixed up to three independent parameters. In section \ref{Section4} we consider three-point 
functions of spinor current and the flavour current multiplets. In section \ref{Section5} we discuss remaining three-point correlators involving just 
the spinor current multiplet. In section \ref{Section6} we study whether our results in sections \ref{Section3} and \ref{Section4} are consistent with $\cN=2$ superconformal symmetry. 
For this we perform superspace reduction from $\cN=2$ to $\cN=1$. We show that our result for the three point function~~\eqref{e00}, 
in general, is not consistent with $\cN=2$ superconformal symmetry. 
Finally, in section \ref{Section7} we discuss the reduction of the superfield correlator~\eqref{e00} to correlators of component currents. 
More precisely, we concentrate on the three-point functions of two $\rm U(1)$ vector currents contained in $S_{\a}$ 
with the axial current and the energy-momentum tensor contained in $J_{\g \gd}$. 
We show that our results are in complete agreement with the general structure of these three-point functions derived in~\cite{OP, EO}.


\section{Superconformal building blocks} \label{Section2}


In this section we provide a brief review of two and three-point superconformal structures in 4D $\cN$-extended superspace, which were introduced in \cite{Park1, OsbornN1} in the ${\cN}=1$ case, and later generalised to arbitrary ${\cN}$ in \cite{Park} (see also \cite{KT} for a review). These superconformal structures are important for constructing correlation functions of conserved current multiplets in later sections. This section closely follows \cite{KT}.


\subsection{Infinitesimal superconformal transformations}


Consider $\cN$-extended Minkowski superspace ${\mathbb M}^{4|4\cN}$ parametrised by $z^A = (x^a, \q^\a_i, {\bar \q}^i_\ad)$, where $a = 0,1, \cdots 3,~ \a, \ad = 1, 2$ and $i = \1, \2, \cdots \underline{\cN}$. The Grassmann variables $\q^\a_i$ and ${\bar \q}^i_\ad$ are related to each other by complex conjugation: $\overline{\q^\a_i} = \bar{\q}^{\ad i}$.
Infinitesimal superconformal transformations 
\be
\d z^A = \xi z^A~,
\ee
are generated by conformal Killing supervector fields \cite{Ideas, KT}
\be
\x = {\overline \x} = \x^a (z) \pa_a + \x^\a_i (z)D^i_\a
+ {\bar \x}_\ad^i (z) {\bar D}^\ad_i
\ee   
which satisfy the equation
\be
[\x \, , \, D^i_\a ] ~\propto ~ D^j_\b ~.
\label{Keqn}
\ee   
Eq. \eqref{Keqn} implies that the spinor parameters can be expressed in terms of the vector ones
\be
\x^\a_i = -\frac{\rm i}{8} {\bar D}_{\bd i} \x^{\bd \a}\;, \qquad
{\bar D}_{\bd j} \x^\a_i = 0~.
\label{spinsc}
\ee
The vector parameters satisfy 
\be
D^i_{(\a} \x_{\b )\bd} = {\bar D}_{i(\ad} \x_{\b \bd )}=0~,
\label{msc}
\ee
which result in the conformal Killing equation
\be
\pa_a \x_b + \pa_b \x_a = \hf\, \eta_{ab}\, \pa_c \x^c\;.
\ee 
The general solution to eq. (\ref{msc}) was given 
in \cite{Ideas} for $\cN=1$ and in \cite{Park} for $\cN >1$. The superalgebra of ${\cN}$-extended conformal Killing supervector fields is isomorphic to $\rm su(2,2|\cN)$.
We have the relation
\be
[\x \;,\; D^i_\a ] = - (D^i_\a \x^\b_j) D^j_\b
= \hat{\o}_\a{}^\b  D^i_\b - \frac{1}{\cN}
\Big( (\cN-2) \s + 2 {\bar \s}  \Big) D^i_\a
-{\rm i} \hat{\L}_j{}^i \; D^j_\a \;.
\ee
Here the superfield parameters are expressed in terms of $\x^A = (\x^a, \x^{\a}_i, \bar \x^i_{\ad})$ as
\be \label{z-dep}
\begin{aligned}
&\hat{\o}_{\a \b}(z) = -\frac{1}{\cN}\;D^i_{(\a} \x_{\b)i}\;,
\qquad \s (z) = \frac{1}{\cN (\cN - 4)}
\left( \hf (\cN-2) D^i_\a \x^\a_i - 
{\bar D}^\ad_i {\bar \x}_{\ad}^{ i} \right)~,\\
&\hat{\L}_j{}^i (z) = -\frac{1}{32}\left(
[D^i_\a\;,{\bar D}_{\ad j}] - \frac{1}{\cN}
\d_j{}^i  [D^k_\a\;,{\bar D}_{\ad k}] \right)\x^{\ad \a}~.
\end{aligned}
\ee
The $z$-dependent parameters $\hat{\o}_{\a \b}(z), \s (z), \hat{\L}_j{}^i (z)$ correspond to the `local' Lorentz, scale and $\rm SU(\cN)$ $R$-symmetry transformations, respectively. One may also check the following properties: 
\be
\begin{aligned}
&{\bar D}_{\ad i} \hat{\o}_{\a \b} = 0\;,
\qquad {\bar D}_{\ad i} \s = 0~, \\
&\hat{\L}^\dag =  \hat{\L}~, \quad  {\rm tr}\; \hat{\L} = 0~, \quad D^k_\a \hat{\L}_j{}^i = 2{\rm i} \left( \d^k_j D^i_\a 
-\frac{1}{\cN} \d^i_j D^k_\a  \right) \s~.
\end{aligned}
\ee
{}For $\cN=2$, the latter property leads to the analyticity condition
\be
D^{(i}_\a \hat{\L}^{jk)} = {\bar D}^{(i}_\ad \hat{\L}^{jk)}= 0~.
\ee
The above formalism cannot be directly applied to the case $\cN = 4$.
In this paper, we are interested in the $\cN = 1$ and $\cN =2$ cases.


\subsection{Two-point structures}


Let $z_1$ and $z_2$ be two different points in superspace. In constructing correlation functions of primary superfields, all building blocks are composed of the following two-point structures
\bsubeq
\bea
x^a_{\bar{1} 2} &=& -x^{a}_{2 \bar{1}} = x^{a}_{1-} - x^a_{2+} + 2 \ri \,\q_{2 i}\s^{a} {\bar \q}^{i}_{1}~,\\
\q_{12} &=& \q_1 - \q_2~, \qquad {\bar \q}_{12} = {\bar \q}_1 - {\bar \q}_2~,
\eea 
\esubeq
where $x^a_{\pm}= x^a \pm \ri \theta_i \sigma^a {\bar \theta}^i$.
The former can also be expressed in spinor notation as
\bsubeq
\bea
({x}_{\bar{1} 2})^{\ad \a} &=& (\tilde{\s_a})^{\ad \a}x^{a}_{\bar{1} 2}~, \\
({x}_{2 \bar{1}})_{\a \ad} &=& (\s_a)_{\a \ad} \, x^a_{2 \bar{1} }=
-(\s_a)_{\a \ad} \, x^a_{\bar{1} 2}= - \ve_{\a \b} \ve_{\ad \bd} ({x}_{\bar{1} 2})^{\bd \b}~,\\
({x}_{\bar{1} 2}^{\,\dagger} )^{\ad \a} &=& - ({x}_{\bar{2} 1})^{\ad \a}~.
\eea
\esubeq
The notation '$x_{\bar{1} 2}$' means that $x_{\bar{1} 2}$ is antichiral with respect to $z_1$ and chiral with respect to $z_2$. 
That is,
\be 
D_{(1)  \a}^i x_{\bar{1} 2}=0\,, \qquad \bar D_{(2) \ad  i} x_{\bar{1} 2}=0~,
\label{e1}
\ee
where $D_{(1) \a}^{i}$ and $\bar D_{(1) \ad i}$  are the superspace covariant spinor derivatives acting on the point $z_1$ and similarly,
$D_{(2) \a}^{i}$ and $\bar D_{(2) \ad i}$ act on the point $z_2$,
\bea
D^{i}_{\a} = \frac{\pa}{\pa \q^{\a}_{i}}+ \ri (\s^a)_{\a \ad} \bar{\q}^{\ad i} \frac{\pa}{\pa x^a}~, \qquad \bar D_{\ad i} = -\frac{\pa}{\pa \bar \q^{\ad i}} - \ri {\q}^{\a}_{i} (\s^a)_{\a \ad}  \frac{\pa}{\pa x^a}~.
\eea
We will also introduce the inverse of $\tilde{x}_{\bar{1} 2}$:
\bea
(\tilde{x}_{\bar{1} 2}{}^{-1})_{\a \ad} = \frac{1}{x_{\bar{1} 2}{}^{2}}( x_{2 \bar{1}})_{\a \ad}~.
\eea
Note that $({x}_{\bar{1} 2})^{\ad \a} (x_{2 \bar{1}})_{\a \bd} = x_{\bar{1} 2}{}^{2}\, \d^{\ad}{}_{\bd}$. 
We will also often make use of matrix-like conventions of \cite{OsbornN1, KT} 
when the spinor indices are not explicitly written. 
We denote 
\bsubeq
\bea
&&
\psi = (\psi^{\a})~, \quad  \tilde{\psi} = (\psi_{\a})~, \quad  \bar{\psi} = (\bar{\psi}^{\ad})~, \quad \tilde{\bar{\psi}} = (\bar{\psi}_{\ad})~, 
\label{e2a}
\\
&&
x = (x_{\a \ad})~, \quad \tilde{x} = (x^{\ad \a})~.
\label{e2b}
\eea
\esubeq
Since $x^2 \equiv x^a x_a = -\hf {\rm tr}(\tilde{x} x)$, it follows that $\tilde{x}^{-1} = -x/x^2$.

The two-point structures transform under the superconformal group as follows
\bsubeq
\bea
\d x^{\ad \a}_{{\bar 1}2} &=& -\Big( \hat{{\bar \o}}^\ad{}_\bd (z_1)   
- \d^\ad{}_\bd \,{\bar \s}(z_1)\Big) x^{\bd \a}_{{\bar 1}2}  
- x^{\ad \b}_{{\bar 1}2} \Big( \hat{\o}_\b{}^\a (z_2) -
\d_\b{}^\a \,\s (z_2) \Big) \label{2ptvar-1} \\
\d \q^\a_{12\,i} &=& {\rm i}  \hat{\L}_i{}^j (z_1)
+ \Big(\frac{2}{\cN}\left({\bar \s}(z_1) - \s (z_1)\right)
\,\d_i{}^j\Big) 
\q^\a_{12\,j} - {\rm i} \,\hat{{\bar \eta}}_{\bd i}(z_1)
x^{\bd \a}_{{\bar 1}2}
\non \\
&& - \q^\b_{12\,i} 
\left( \hat{\o}_\b{}^\a (z_2) -
\d_\b{}^\a \,\s (z_2) \right)~, 
\label{two-point-variation} 
\eea 
\esubeq
with $\hat{\eta}_\a{}^i (z) := \hf D^i_\a \s (z)$.
It is useful to define
\bea
(\hat{x}_{2 \bar{1}})_{\a \ad} = \frac{(x_{2 \bar{1}})_{\a \ad}}{(x_{\bar{1} 2}{}^2)^\hf} \in {\rm SL(2, \mathbb{C})}~, \qquad (\hat{x}_{1 \bar{2}})_{\a \ad} = -(\hat{x}_{2 \bar{1} }^{\,\dagger} )_{\a \ad}~.
\eea 
From \eqref{2ptvar-1}, it follows that $x_{\bar{1} 2}{}^{2}$ and $(\hat{x}_{2 \bar{1}})_{\a \ad}$ transform covariantly under superconformal 
transformations:
\bsubeq
\bea
\d x_{\bar{1} 2}{}^{2} &=& 2 \big( \bar{\s}(z_1) + \s(z_2) \big)x_{\bar{1} 2}{}^{2}~, \\
\d (\hat{x}_{2 \bar{1}})_{\a \ad} &=& (\hat{x}_{2 \bar{1}})_{\a \gd} \, \hat{\bar \o}^{\gd}{}_{\ad} (z_1) + \hat{\o}^{\g}{}_{\a} (z_2)\, (\hat{x}_{2 \bar{1}})_{\g \ad}~. \label{xhat}
\eea
\esubeq
Hence, they appear as building blocks in correlation functions of primary operators. 

Another important structure is the conformally covariant ${\cN} \times \cN$ matrix
\bea
u_i\,^j (z_{12}) = \d_i\,^j + 4 \ri \frac{\q_{12 i} \, {x}_{2 \bar{1}} \,\bar{\q}^j_{12}}{x_{\bar{1} 2}{}^2} = \d_i\,^j + 4 \ri \q_{12 i} \, \tilde{x}_{ \bar{1} 2}{}^{-1} \,\bar{\q}^j_{12}~,
\eea
which satisfies
\bea
u^{\dagger}(z_{12}) u(z_{12}) = 1~, \qquad u^{-1} (z_{12}) = u(z_{21})~, \qquad {\rm det}\, u(z_{12}) = \frac{x_{\bar{1} 2}{}^2}{x_{\bar{2} 1}{}^2}~.
\eea
The unimodular unitary matrix
\bea
\hat{u}_{i}\,^{j} (z_{12}) = \Big( \frac{x_{\bar{2} 1}{}^2}{x_{\bar{1} 2}{}^2} \Big)^{1/{\cN}} u_{i}\,^{j} (z_{12})
\eea
has the transformation rule
\bea
\d \hat{u}_{i}\,^{j} (z_{12}) = \ri \hat{\Lambda}_{i}\,^{k}(z_1) \hat{u}_{k}\,^{j} (z_{12}) - \ri \hat{u}_{i}\,^{k} (z_{12}) \hat{\Lambda}_{k}\,^{j}(z_2)~.
\label{uij-trf}
\eea
As follows from \eqref{xhat} and \eqref{uij-trf}, the two-point structures $(\hat{x}_{2 \bar{1}})_{\a \ad}$ and $\hat{u}_{i}{}^{j} (z_{12})$ transform covariantly with respect to the superconformal group, i.e., as tensors with Lorentz and $\rm SU(\cN)$ indices at both superspace points.

In the ${\cN}=2$ case, using the $\rm SU(2)$-invariant tensors $\ve_{ij} = -\ve_{ji}$ and $\ve^{ij} = - \ve^{ji}$, with $\ve^{\1 \2} = \ve_{\2 \1} = 1$, we can raise and lower isoindices 
\bea
C^i = \ve^{ij} C_j~, \qquad C_i = \ve_{ij} C^j~. 
\eea
We can then write the condition of unimodularity of the matrix $\hat{u}$ as
\bea
(\hat{u}^{-1} (z_{12}))_{i}\,^{j} = \hat{u}_{i}\,^{j} (z_{21}) = \ve^{jk} \hat{u}_{k}\,^{l} (z_{12})\ve_{li}~,
\eea
or, equivalently
\bea
\hat{u}_{ji} (z_{21}) = -\hat{u}_{ij} (z_{12})~.
\eea

To finish this subsection let us also note several useful differential identities:
\bsubeq
\bea
&&D_{(1) \a}^{i} (x_{\bar{2} 1})^{\bd \b} = 4 \ri \d_{\a}{}^{\b} \bar{\q}_{12}^{\bd i}~, \qquad \bar{D}_{(1) \ad\, i}\, (x_{\bar{1} 2})^{\bd \b} = 4 \ri \d_{\ad}{}^{\bd} \q_{12 i}^{\b}~, \\
&&D_{(1) \a}^{i} \bigg(\frac{1}{x_{\bar{2} 1}{}^{2}} \bigg) = - \frac{4 \ri}{x_{\bar{2} 1}{}^2}  (\tilde{x}_{\bar{2} 1}{}^{-1})_{\a \bd} \bar{\q}_{12}^{\bd \,i}~, \quad \bar D_{(1) \ad \,i} \bigg(\frac{1}{x_{\bar{1} 2}{}^{2}} \bigg) = - \frac{4 \ri}{x_{\bar{1} 2}{}^2}  (\tilde{x}_{\bar{1} 2}{}^{-1})_{\a \ad} {\q}_{12\, i}^{\a}~,~~~~~
\eea
\esubeq
Here and throughout the paper we assume that the space points are not coincident, $x_1 \neq x_2$. 


\subsection{Three-point structures}


Associated with three superspace points $z_1, z_2$ and $z_3$ are the following superconformally covariant bosonic and fermionic 
variables $\bm{Z}_1, \bm{Z}_2$ and $\bm{Z}_3$, where $\bm{Z}_1 = (\bm{X}_1^a, \bm{\Q}_1^{\a i}, \bm{\bar \Q}_{1i}^{\ad})$ 
(see \cite{OsbornN1, Park1, KT} for details):
\bsubeq
\bea
&&\bm{X}_{1} = \tilde{x}^{-1}_{\bar{2}1} \tilde{x}_{\bar{2} 3} \tilde{x}^{-1}_{\bar{1} 3}~, \,\,\, \tilde{\bm{\Q}}_{1}^i = \ri \big(\tilde{x}^{-1}_{\bar{2} 1} \bar{\q}_{12}^i -  \tilde{x}^{-1}_{\bar{3} 1} \bar{\q}_{13}^i\big)~, \,\,\, \tilde{\bar{\bm{\Q}}}_{1 i} = \ri \big({\q}_{12 i} \tilde{x}^{-1}_{\bar{1} 2} \ - {\q}_{13 i} \tilde{x}^{-1}_{\bar{1} 3} \big)~,~~~~~ \label{Z1}\\
&&\bm{X}_{2} = \tilde{x}^{-1}_{\bar{3}2} \tilde{x}_{\bar{3} 1} \tilde{x}^{-1}_{\bar{2} 1}~, \,\,\, \tilde{\bm{\Q}}_{2}^i = \ri \big(\tilde{x}^{-1}_{\bar{3} 2} \bar{\q}_{23}^i -  \tilde{x}^{-1}_{\bar{1} 2} \bar{\q}_{21}^i\big)~, \,\,\, \tilde{\bar{\bm{\Q}}}_{2 i} = \ri \big({\q}_{23 i} \tilde{x}^{-1}_{\bar{2} 3} \ - {\q}_{21 i} \tilde{x}^{-1}_{\bar{2} 1} \big)~,~~~~~ \label{Z2}\\
&&\bm{X}_{3} = \tilde{x}^{-1}_{\bar{1}3} \tilde{x}_{\bar{1} 2} \tilde{x}^{-1}_{\bar{3} 2}~, \,\,\, \tilde{\bm{\Q}}_{3}^i = \ri \big(\tilde{x}^{-1}_{\bar{1} 3} \bar{\q}_{31}^i -  \tilde{x}^{-1}_{\bar{2} 3} \bar{\q}_{32}^i\big)~, \,\,\, \tilde{\bar{\bm{\Q}}}_{3 i} = \ri \big({\q}_{31 i} \tilde{x}^{-1}_{\bar{3} 1} \ - {\q}_{32 i} \tilde{x}^{-1}_{\bar{3} 2} \big)~.~~~~~ \label{Z3}
\eea
\esubeq
Clearly, the structures in \eqref{Z2} and \eqref{Z3} are obtained through cyclic permutations of superspace points. Thus, it suffices to study the properties of \eqref{Z1}. We also define
\bea
\bar{\bm{X}}_{1} = \bm{X}_1^{\dagger} &=& -\tilde{x}_{\bar{3}1}{}^{-1} \tilde{x}_{\bar{3} 2} \tilde{x}_{\bar{1} 2}{}^{-1}~,
\eea
and similar relations hold for $\bar{\bm{X}}_{2}, \bar{\bm{X}}_{3}$. The structures \eqref{Z1} transform as tensors at $z_1$ 
\bsubeq \label{Z1transf}
\bea
\d \bm{X}_{1\a \ad} &=& \Big( \hat{\o}_{\a}\,^{\b}(z_1) - \d_{\a}\,^{\b}\s(z_1)\Big)  \bm{X}_{1\b \ad} +  \bm{X}_{1\a \bd} \Big( \hat{\bar{\o}}^{\bd}\,_{\ad} (z_1) - \d^{\bd}\,_{\ad} \bar{\s}(z_1)\Big)~,\\
\d {\bm \Q}^i_{1\, \a} & = & 
\hat{\o}_\a{}^\b (z_1) {\bm \Q}^i_{1\, \b} -
{\rm i} {\bm \Q}^j_{1\, \a} \hat{\L}_j{}^i (z_1)
-\frac{1}{\cN} \Big( (\cN - 2) \s (z_1) + 
2{\bar \s} (z_1) \Big){\bm \Q}^i_{1\, \a} ~.
\eea
\esubeq 
By cyclic permutation of labels, we may see that $\bm{Z}_i$ transform as tensors at the point $z_i,\, i = 1,2,3$.

Using the matrices $u (z_{rs})$, $r,s = 1,2,3$, it is possible to construct unitary matrices
\bea
{\bf u} (\bZ_3) &=& u (z_{31}) u (z_{12})u (z_{23})~, \qquad
{\bf u}_i{}^j (\bZ_3) = \d_i{}^j 
- 4{\rm i} \tilde{ \bar{\bm{\Q} }} _{3\,i} {\bm X}_3{}^{-1} \tilde{\bm{\Q}}^j_3 \non \\
{\bf u}^\dag (\bZ_3) &=& u (z_{32}) u (z_{21})u (z_{13})~, \qquad
{\bf u}^\dag_i{}^j(\bZ_3)=  \d_i{}^j + 4{\rm i}
\tilde{ \bar{\bm{\Q}} }_{3\,i} \bar {\bm X}_3{}^{-1} 
\tilde{\bm{\Q}}^j_3
\label{greatu}
\eea
transforming covariantly at $z_3$ only. Their properties are
\be
{\bf u}^\dag (\bZ_3)= {\bf u}^{-1} (\bZ_3)~,\qquad
\qquad 
\det  {\bf u} (\bZ_3) = \frac{ {\bm X}_3^2 }
{ \bar {\bm X}_3^2 }~.
\ee

In the $\cN=1$ case, there are several properties of $\bm{Z}$'s which might be useful later. 
\bsubeq
\bea
&&\bm{X}_1^2 = \frac{x_{\bar{2} 3}{}^2}{x_{\bar{2} 1}{}^2 x_{\bar{1} 3}{}^2}~, \qquad \bm{\bar X}_1^2 = \frac{x_{\bar{3} 2}{}^2}{x_{\bar{3} 1}{}^2 x_{\bar{1} 2}{}^2}~, \\
&&\bm{\bar X}_{1 \a \ad} = \bm{X}_{1 \a \ad} + \ri \bm{P}_{1 \a \ad}~, \quad \bm{P}_{1 \a \ad} = -4 \bm{\Q}_{1\a} \bm{\bar \Q}_{1\ad}~, \quad \bm{P}_1^2 = -8 \bm{\Q}_1^2 \bm{\bar \Q}_1^2~, \\
&&\frac{1}{\bm{\bar X}_1^2} = \frac{1}{\bm{X}_1^2} -2\ri \frac{(\bm{P}_1 \cdot \bm{X}_1)}{(\bm{X}_1^2)^2}~, \quad (\bm{P}_1 \cdot \bm{X}_1) = -\hf \bm{P}_1^{\ad \a} \bm{X}_{1 \a \ad}~. 
\eea
\esubeq
Note that $\bm{\bar X}$ is not an independent variable since it can be expressed in terms of 
$\bm{X}, \bm{\Q}, \bm{\bar \Q}$.
The variables $\bm{Z}$ with different labels are related to each other via the identities
\bsubeq \label{Z13}
\bea
\tilde{x}_{\bar{1} 3} \bm{X}_3 \tilde{x}_{\bar{3} 1} &=& - \bar{\bm{X}}_1^{-1} = \frac{\tilde {\bar{ \bm{X}}}_1}{\bar {\bm{X}}_1^2}~, \qquad \tilde{x}_{\bar{1} 3} \bar{\bm{X}}_3 \tilde{x}_{\bar{3} 1} = - \bm{X}_1^{-1} = \frac{\tilde { \bm{X}}_1}{\bm{X}_1^2}~, \\
\frac{x_{\bar{3} 1}{}^2}{x_{\bar{1} 3}{}^2} \tilde{x}_{\bar{1} 3} \tilde{\bm{\Q}}_3 &=&
 - \bm{X}_1^{-1} \tilde{\bm{\Q}}_1~, \qquad \frac{x_{\bar{1} 3}{}^2}{x_{\bar{3} 1}{}^2}  \tilde{\bar {\bm{\Q}}}_3 \tilde{x}_{\bar{3} 1} =
  \tilde{\bar{\bm {\Q}}}_1 \bar{\bm{X}}_1^{-1} ~.
\eea
\esubeq
Making use of \eqref{Z1transf} and \eqref{Z13}, it follows that
\bea
\frac{\bm{X}_1^2}{\bm{\bar X}_1^2} = \frac{\bm{X}_2^2}{\bm{\bar X}_2^2} = \frac{\bm{X}_3^2}{\bm{\bar X}_3^2}~,
\eea
and this combination is a superconformal invariant. 


\subsection{Correlation functions of primary superfields}


Consider a tensor superfield $\cO^{\cA}_{\cI}(z)$ transforming in a representation $T$ of the Lorentz group with respect to the index $\cA$, 
and in the representation $D$ of the $R$-symmetry group SU$(\cN)$ with respect to the index 
$\cI$.\footnote{We assume the representations $T$ and $D$ are irreducible.} Such a superfield is called primary if its infinitesimal superconformal transformation law reads
\be
\begin{aligned}
\d\, \cO^\cA_\cI (z) &= - \x \, \cO^\cA_\cI (z) 
+ (\hat{\o}^{\a \b} (z) M_{\a \b}+ 
\hat{ \bar{\o}}^{\dot{\a} \dot{\b}} (z) 
\bar{M}_{\dot{\a} \dot{\b}} )^\cA{}_\cB\,
\cO^\cB_\cI (z) \\
& + {\rm i} \, \hat{\L}^k{}_l (z)\,
(R^l{}_k )_\cI{}^\cJ \cO^\cA_\cJ (z) 
 - 2\left( q\, \s(z) + \bar{q}\, \bar{\s} (z) \right) 
\cO^\cA_\cI (z)~.
\end{aligned}
\ee
In the above, $\xi$ is the superconformal Killing vector, while $\hat{\o}^{\a \b} (z), ~\s(z), ~\hat{\L}^k{}_l (z)$ are the $z$-dependent parameters associated with $\xi$, see eq. \eqref{z-dep}. The superscript `$\cA$' collectively denotes the undotted and dotted spinor indices on which the Lorentz generators $M_{\a \b}$ and $\bar{M}_{\ad \bd}$ act. The matrices $R^i{}_j$ are the SU$(\cN)$ generators.
The weights $q$ and $\bar q$ determine the dimension  $(q+\bar q)$
and U(1) $R$-symmetry charge $(q-\bar q)$ of the superfield.

Various primary superfields, including conserved current multiplets, obey certain differential constraints imposed by their conservation equations. These constraints need to be taken into account when computing correlation functions. 
It proves beneficial to make use of these conformally covariant 
operators \cite{OsbornN1, KT}: $\cD_{\bar A} = (\pa / \pa \bX^a_3, \cD_{\a i}, \bar{\cD}^{\ad i}) $
and $\cQ_{\bar A} = (\pa / \pa \bX^a_3, {\cal Q}_{\a i},
\bar{\cal Q}^{\ad i})$ defined by 
\be
\begin{aligned}
&\cD_{\a i} = \frac{\pa}{ \pa {\bm \Q}^{\a i}_3 }
-2{\rm i}\,(\s^a)_{\a \ad} \bar{\bm \Q}^\ad_{3\, i}
\frac{\pa }{ \pa \bX^a_3 }~, \qquad
\bar{\cD}^{\ad i} = \frac{\pa}{ \pa \bar{\bm \Q}_{3\, \ad i} }~, \\
&{\cal Q}_{\a i}  =  \frac{\pa}{ \pa {\bm \Q}^{\a i}_3 }~, \qquad
\bar{\cal Q}^{\ad i} = \frac{\pa}{ \pa \bar{\bm \Q}_{3\, \ad i} } 
+ 2{\rm i}\, {\bm \Q}^i_{3\, \a} (\tilde{\s}^a)^{\ad \a}
\frac{\pa}{ \pa \bX^a_3 }~,  \\
& [\cD_{\bar A}  ,  \cQ_{\bar B} \} ~=~0 ~. 
\end{aligned}
\ee
They satisfy anti-commutation relations
\bea
\{ \cD^{\a}_{i}, \bar{\cD}^{\ad j} \} = 2 \ri\, \d_{i}{}^{j} (\tilde{\s}^a)^{\ad \a} \frac{\pa}{\pa \bm X^a}~, \qquad \{ {\cQ}^{\a}_{i}, \bar{\cQ}^{\ad j} \} = -2 \ri\, \d_{i}{}^{j} (\tilde{\s}^a)^{\ad \a} \frac{\pa}{\pa \bm X^a}~.
\eea

Given an arbitrary function $t (\fn3)$, the following differential identities hold
\bsubeq \label{N2covderiv}
\bea
D^{i}_{(1)\a} \,t(\fn3) &=& -\frac{\ri}{x_{\bar{3} 1}{}^2}(x_{1 \bar{3}})_{\a \ad} \, u_{j}\,^{i} (z_{31})\bar{\cD}^{\ad\, j} t(\fn3)~,\\
\bar{D}_{(1)\ad\, i} \,t(\fn3) &=& -\frac{\ri}{x_{\bar{1} 3}{}^2}(x_{3 \bar{1}})_{\a \ad} \,u_{i}\,^{j} (z_{13}) {\cD}^{\a}_{j} t(\fn3)~,\\
D^{i}_{(2)\a}\, t(\fn3) &=& -\frac{\ri}{x_{\bar{3} 2}{}^2}(x_{2 \bar{3}})_{\a \ad} \,u_{j}\,^{i} (z_{32})\bar{\cQ}^{\ad\, j} t(\fn3)~,\\
\bar{D}_{(2) \ad\, i}\, t(\fn3) &=& -\frac{\ri}{x_{\bar{2} 3}{}^2}(x_{3 \bar{2}})_{\a \ad}\, u_{i}\,^{j} (z_{23}) {\cQ}^{\a}_{j} t(\fn3)~.
\eea
\esubeq
The above identities can be derived using these relations:
\bsubeq
\bea
&&\bar D_{(1)\ad i} \bm X_{3 \g \gd} = -4 (\tilde{x}^{-1}_{\bar{1} 3})_{\g \ad} \,u_{i}{}^{j} (z_{13}) \bar{\bm \Q}_{3 \gd j}~,\quad 
\bar D_{(1)\ad i} \,\bm \Q_{3 \g}^j = \ri (\tilde{x}^{-1}_{\bar{1} 3})_{\g \ad} \,u_{i}{}^{j} (z_{13})~,\\
&&\bar D_{(1)\ad i} \bar {\bm X}_{3 \g \gd} = 0~, \qquad \bar D_{(1)\ad i} \bar{\bm \Q}_{3 \gd j}= 0~, \\
&&D_{(2) \b}^{i} \bm X_{3 \g \gd} = 4 (\tilde{x}^{-1}_{\bar{3} 2})_{\b \gd} {\bm \Q}_{3 \g}^{j} \,u_{j}{}^{i} (z_{32}) ~,\quad D_{(2) \b}^{i} \,\bar{\bm \Q}_{3 \gd j} = \ri (\tilde{x}^{-1}_{\bar{3} 2})_{\b \gd} \,u_{j}{}^{i} (z_{32}) ~,\\
&&D_{(2)\b}^{i} \bar{\bm X}_{3 \g \gd} = 0~, \qquad D_{(2)\b}^{i} {\bm \Q}_{3 \g}^{j} = 0~.
\eea
\esubeq
In the ${\cN}=1$ case, we arrive at the following
\bsubeq \label{Cderivs}
\bea
D_{(1)\a} \,t(\fn3) &=& -\frac{\ri}{x_{\bar{1} 3}{}^2}(x_{1 \bar{3}})_{\a \ad} \bar{\cD}^{\ad} t(\fn3)~,\\
\bar{D}_{(1)\ad} \,t(\fn3) &=& -\frac{\ri}{x_{\bar{3} 1}{}^2}(x_{3 \bar{1}})_{\a \ad} {\cD}^{\a} t(\fn3)~,\\
D_{(2)\a}\, t(\fn3) &=& \frac{\ri}{x_{\bar{2} 3}{}^2}(x_{2 \bar{3}})_{\a \ad} \bar{\cQ}^{\ad} t(\fn3)~,\\
\bar{D}_{(2) \ad}\, t(\fn3) &=& \frac{\ri}{x_{\bar{3} 2}{}^2}(x_{3 \bar{2}})_{\a \ad} {\cQ}^{\a} t(\fn3)~.
\eea
\esubeq

In accordance with the general prescription of \cite{OsbornN1,Park1, Park, KT},
the two-point correlation function of a primary superfield
$\cO^{\cA}_\cI$ with its
conjugate $\bar{\cO}^\cJ_{\cB}$ takes the form
\be
\langle \cO^{\cA}_\cI (z_1)\;\bar{\cO}_{\cB}^\cJ (z_2)\rangle
~=~ C_{\cO}\;\frac{ 
T^{\cA}{}_{\cB} ({\hat{x}_{1 \bar{2}}}, {\hat{x}_{2 \bar{1}}})\,  D_\cI{}^\cJ \left( \hat{u}(z_{12}) \right)}
{ (x_{\bar{1}2}{}^2)^{\bar q} (x_{\bar{2}1}{}^2)^q }~,
\ee
with $ C_{\cO}$ being a normalisation constant. 
Furthermore, the three-point correlation function 
of primary superfields $\F_{\cI_1}^{\cA_1}, \J_{\cI_2}^{\cA_2}$ and $\Pi_{\cI_3}^{\cA_3}$ has the general expression \cite{OsbornN1, Park1, Park, KT}:
\bea \label{3ptgen}
&&\langle
\F_{\cI_1}^{\cA_1} (z_1) \, \J_{\cI_2}^{\cA_2}(z_2)\,  \Pi_{\cI_3}^{\cA_3}(z_3)
\rangle  \\
&&\quad =\frac{ 
T^{(1)\cA_1}{}_{\cB_1} ({\hat{x}_{1 \bar{3}}}, {\hat{x}_{3 \bar{1}}})\,
T^{(2)\cA_2}{}_{\cB_2} ({\hat{x}_{2 \bar{3}}}, {\hat{x}_{3 \bar{2}}})\,
D^{(1)}{}_{\cI_1}{}^{\cJ_1} \left( \hat{u}(z_{13}) \right)
D^{(2)}{}_{\cI_2}{}^{\cJ_2} \left( \hat{u}(z_{23}) \right)
}
{ 
(x_{\bar{1}3}{}^2)^{\bar{q}_1} (x_{\bar{3}1}{}^2)^{q_1} 
(x_{\bar{2}3}{}^2)^{\bar{q}_2} (x_{\bar{3}2}{}^2)^{q_2}
} \non\\
&&\qquad \qquad \times 
H^{\cB_1 \cB_2 \cA_3}_{\cJ_1 \cJ_2 \cI_3} (\fn3)~.\non 
\eea 
The functional form of the tensor $H^{\cB_1 \cB_2 \cA_3}_{\cJ_1 \cJ_2 \cI_3}$ is highly constrained by the superconformal symmetry as follows:
\begin{itemize}

\item[(i)] It possesses the homogeneity property
\be
\begin{aligned}
& H^{\cB_1 \cB_2 \cA_3}_{\cJ_1 \cJ_2 \cI_3} ( \l \bar{\l}\, {\bm X},
\l\, {\bm \Q}, \bar{\l} \bar {\bm \Q}) =
\l^{2a} \bar{\l}^{2\bar{a}}
H^{\cB_1 \cB_2 \cA_3}_{\cJ_1 \cJ_2 \cI_3} ( {\bm X}, {\bm \Q}, \bar {\bm \Q} ) \\
& \Big( \frac{2}{\cN}-1\Big)a- \frac{2}{\cN}\bar{a} = \bar{q}_1 + \bar{q}_2 -q_3~,\qquad
\Big( \frac{2}{\cN}-1\Big)\bar{a} - \frac{2}{\cN}a = q_1 + q_2 - \bar{q}_3~,
\end{aligned}
\ee
which guarantees that the correlation function has the correct transformation law under the superconformal group.
Note that eq.~\eqref{3ptgen} by construction has the correct transformation properties at the points $z_1$ and $z_2$. 
The above homogeneity property implies that it transforms correctly also at the point $z_3$.

\item[(ii)]If any of the superfields $\F$, $\J$ and $\P$ satisfy differential equations (e.g. conservation laws of conserved current multiplets), then the tensor
$H^{{\cal J}_1 {\cal J}_2 {\cal I}_3}_{{\cal B}_1 {\cal B}_2 {\cal A}_3}$
is constrained by certain differential equations as well. The latter may be derived using \eqref{N2covderiv}.

\item[(iii)] If any (or all) of
the superfields $\F$, $\J$ and $\P$ coincide,
the tensor $H^{{\cal J}_1 {\cal J}_2 {\cal I}_3}_{{\cal B}_1 {\cal B}_2 {\cal A}_3}$
obeys certain constraints, as a consequence of the symmetry
under permutations of superspace points, e.g.
\be
\langle \Phi_{{\cal I}}^{{\cal A}}(z_1) \Phi_{{\cal J}}^{{\cal B}}(z_2)
\P_{{\cal K}}^{{\cal C}}(z_3) \rangle =
(-1)^{\epsilon(\Phi)}
\langle \Phi_{{\cal J}}^{{\cal B}}(z_2) \Phi_{{\cal I}}^{{\cal A}}(z_1)
\P_{{\cal K}}^{{\cal C}}(z_3) \rangle~,
\ee
where $\epsilon(\Phi)$ is the Grassmann parity of $\Phi_{{\cal I}}^{{\cal A}}$. We further note that under permutations of any two superspace points, the three-point building blocks transform as
\bsubeq
\bea
			{\bm X}_{3 \, \a \ad} &\stackrel{1 \leftrightarrow 2}{\longrightarrow} - \bar{\bm X}_{3 \, \a \ad} \, , \hspace{10mm} {\bm \Q}^{i}_{3 \, \a} \stackrel{1 \leftrightarrow 2}{\longrightarrow} - {\bm \Q}^{i}_{3 \, \a} \, , \label{pt12} \\[2mm]
			{\bm X}_{3 \, \a \ad} &\stackrel{2 \leftrightarrow 3}{\longrightarrow} - \bar{\bm X}_{2 \, \a \ad} \, , \hspace{10mm} {\bm \Q}^{i}_{3 \, \a} \stackrel{2 \leftrightarrow 3}{\longrightarrow} - {\bm \Q}^{i}_{2 \, \a} \, , \label{pt23} \\[2mm]
			{\bm X}_{3 \, \a \ad} &\stackrel{1 \leftrightarrow 3}{\longrightarrow} - \bar{\bm X}_{1 \, \a \ad} \, , \hspace{10mm} {\bm \Q}^{i}_{3 \, \a} \stackrel{1 \leftrightarrow 3}{\longrightarrow} - {\bm \Q}^{i}_{1 \, \a} \, . \label{pt13}
\eea
\esubeq
\end{itemize}
The above conditions fix the functional form of $H^{{\cal J}_1 {\cal J}_2 {\cal I}_3}_{{\cal B}_1 {\cal B}_2 {\cal A}_3}$ (and therefore the three-point correlation function) up to a few arbitrary constants.

The main objective of this paper is to determine the constraints on the general structure of three-point correlation functions  involving a conserved spinor current multiplet $S_{\a}(z)$ (and its conjugate $\bar S_{\ad}(z)$), imposed by ${\cN}=1$ superconformal symmetry.
The spinor current superfield $S_{\a}$ obeys the following constraints
\bsubeq \label{c1}
\bea
D^{\a}S_{\a} &=& 0 \quad \Longrightarrow \quad D^2 S_{\a} = 0~, \\
\bar D^2 S_{\a} &=& 0~.
\eea
\esubeq
These conditions fix its superconformal transformation law to be
\be
\d S_{\a} = -\xi S_{\a} + \hat{\o}_{\a}\,^{\b}S_{\b} - (3 \s + 2\bar{\s}) S_{\a}~,
\ee
and hence, $S_{\a}$ is a primary superfield with weights $(q, \bar q)= (\frac{3}{2}, 1)$ and
dimension $\frac{5}{2}$. 

It was first demonstrated in \cite{KT} (and also discussed in the introduction) that the spinor current multiplet $S_{\a}$ naturally arises from the 
reduction of the ${\cN}=2$ conformal supercurrent \cite{Sohnius:1978pk} to ${\cN}=1$ superspace.\footnote{ 
In a more general setting, Ref. \cite{KR} describes the structure of conformal current supermultiplets of arbitrary ranks in a supergravity background.} 
However, in this paper we will not assume ${\cN}=2$  superconformal symmetry. We will consider ${\cN}=1$ superconformal 
theory which also has a conserved spinor current multiplet $S_{\a}$. Our aim is to find how three-point correlation functions involving $S_{\a}$
are determined by ${\cN}=1$  superconformal symmetry. A surprising result is that these correlation functions, in general, 
do not imply ${\cN}=2$ superconformal symmetry. 


\section{Correlation functions of spinor current multiplets with the supercurrent}  \label{Section3}


Let us recall that the $\cN=1$ supercurrent is described by a primary real superfield $J_{\a \ad} = \bar J_{\a \ad}$, subject to the conservation law
\bea
D^{\a} J_{\a \ad} = \bar{D}^{\ad} J_{\a \ad} = 0~. \label{c2}
\eea
Its superconformal transformation is 
\be
\d J_{\a \ad} = -\xi J_{\a \ad} + (\hat{\o}_{\a}\,^{\b} \d_{\ad}^{\,\bd} + \bar{\hat{\o}}_{\ad}\,^{\bd} \d_{\a}^{\,\b}) J_{\b \bd} -3(\s + \bar{\s}) J_{\a \ad}~.
\ee
The reality constraint and conservation law \eqref{c2} imply that $J_{\a \ad}$ has weights $(q, \bar q)= (\frac{3}{2}, \frac{3}{2})$ and dimension $3$.


\subsection{Correlator $\la \bar S_{\ad}(z_1) S_{\b}(z_2) J_{\g \gd}(z_3) \ra$} \label{Subsec3.1}


According to the general prescription \eqref{3ptgen}, we shall look for the three-point function of the form
\be
\begin{aligned}
\la \bar S_{\ad}(z_1) S_{\b}(z_2) J_{\g \gd}(z_3) \ra 
&= \frac{(x_{3 \bar 1})_{\a \ad} (x_{2 \bar 3})_{\b \bd}} {(x_{\bar 1 3}{}^2)^{2} (x_{\bar 3 2}{}^2)^{2} x_{\bar 3 1}{}^2 x_{\bar 2 3}{}^2} 
H^{\bd \a}\,_{\g \gd} (\fn3) ~. \label{corr1}
\end{aligned}
\ee
The tensor $H_{\a \bd, \, \g \gd}$ has the following homogeneity property
\bea
H_{\a \bd,}\,_{\g \gd}(\l \bar{\l} \bm{X}, \l \bm{\Q}, \bar{\l} \bar{\bm \Q} ) = \l^{-2} \bar \l^{-2} H_{\a \bd, \, \g \gd} (\bm{X}, \bm{\Q}, \bar{\bm \Q})~. \label{sc1}
\eea

The correlator must obey the conservation equations \eqref{c1} and \eqref{c2}. Specifically, we impose:
\bsubeq
\bea
\bar D^{\ad}_{(1)}  \corr1 &=& 0~, \qquad D^2_{(1)}  \corr1 = 0~,\\
D^{\b}_{(2)}  \corr1 &=& 0~, \qquad \bar D^2_{(2)}  \corr1 = 0~,\\
D^{\g}_{(3)}  \corr1 &=& 0~,\qquad
\bar D^{\gd}_{(3)}  \corr1 = 0~. \label{ceq3rd}
\eea
\esubeq
Upon the use of identities \eqref{Cderivs}, along with
\bea
D^{\a}_{(1)} \bigg[ \frac{(x_{1 \bar{3}})_{\a \ad}}{(x_{\bar{3} 1}{}^2)^2}\bigg] = 0 \quad \Longleftrightarrow \quad {\bar D}^{\ad}_{(1)} \bigg[ \frac{(x_{3 \bar{1}})_{\a \ad}}{(x_{\bar{1} 3}{}^2)^2}\bigg] = 0~,
\eea
the constraints on the superspace points $z_1$ and $z_2$ are translated to 
\bsubeq \label{ceq}
\bea
\cD^{\a}H_{\a\bd, \g \gd} = 0~, \label{ceq1}\\
\bar \cD^2 H_{\a\bd, \g \gd} = 0~, \label{ceq2}\\
\bar \cQ^{\bd}H_{\a\bd, \g \gd} = 0~, \label{ceq3}\\
\cQ^2 H_{\a\bd, \g \gd} = 0~. \label{ceq4}
\eea
\esubeq
Imposing differential constraints \eqref{ceq3rd} is more complicated. We will take care of \eqref{ceq3rd} at the end.  

Our aim is to solve for $H_{\a\bd, \g \gd}$ subject to the constraints \eqref{sc1} and \eqref{ceq}.
First, noting that $H_{\a\bd, \g \gd}$ is Grassmann even, the most general ansatz takes the form 
\bea
H_{\a \bd, \g \gd} (\fxq) &=& F_{\a \bd, \g \gd} (\bm{X}) + A_{\a \bd, \g \gd} (\bm{X})\bm{\Q}^2 + B_{\a \bd, \g \gd} (\bm{X}) \bar{\bm \Q}^2 + C_{\a \bd, \g \gd}\bm{\Q}^2 \bar{\bm \Q}^2 \non\\
&&+ G_{\a \bd, \g \gd, \d \dot{\d}} (\bm{X}) \bm{\Q}^{\d} \bar{\bm \Q}^{\dot{\d}}~. 
\eea
It is easy to see that the conditions \eqref{ceq2} and \eqref{ceq4} lead to
\bea
A_{\a \bd, \,\g \gd} (\bm{X}) = 0~,\quad B_{\a \bd, \,\g \gd} (\bm{X}) = 0~, \quad C_{\a \bd, \,\g \gd} (\bm{X}) = 0~,
\eea
and hence, $H_{\a \bd, \g \gd} (\bm{X}, \bm{\Q}, \bm{\bar{\Q}})$ gets simplified to
\bea
H_{\a \bd, \g \gd} (\bm{X}, \bm{\Q}, \bm{\bar{\Q}}) &=& F_{\a \bd, \g \gd} (\bm{X}) + G_{\a \bd, \g \gd, \d \dot{\d}} (\bm{X}) \bm{\Q}^{\d} \bm{\bar \Q}^{\dot{\d}}~. \label{Hfg}
\eea
In order to work out the structures of $F_{\a \bd, \g \gd} (\bm{X})$ and $G_{\a \bd, \g \gd, \d \dot{\d}} (\bm{X})$, it is instructive to make use of vector notation, where we write
\bsubeq
\bea
F_{\a \bd,\,  \g \gd} &=& (\s^m)_{\a \bd} (\s^n)_{\g \gd} F_{mn}~,\\
G_{\a \bd, \,\g \gd, \,\d \dot{\d}} &=& (\s^m)_{\a \bd} (\s^n)_{\g \gd} (\s^p)_{\d \dot{\d}} G_{mnp}~.
\eea
\esubeq
It is not difficult to construct 
the most general expressions for these tensors, which are consistent with eq. \eqref{sc1}. We get
\bea
&&F_{mn} = c_1 \frac{\eta_{mn}}{\bm{X}^2} + c_2 \frac{\bm{X}_m \bm{X}_n}{(\bm{X}^2)^2}~, \\
&&G_{mnp} = \ri d_1 \frac{\bm{X}_m \bm{X}_n \bm{X}_p}{(\bm{X}^2)^3} + \ri d_2 \frac{\eta_{mn} \bm{X}_p}{(\bm{X}^2)^2} +  \ri d_3 \frac{\eta_{mp} \bm{X}_n}{(\bm{X}^2)^2} +  \ri d_4 \frac{\eta_{np} \bm{X}_m}{(\bm{X}^2)^2} + d_5 \e_{mnpq}\frac{\bm{X}^q}{(\bm{X}^2)^2}~,~~~~~~
\eea
where $c_n$ and $d_n$ are some complex coefficients. Our next task is to rewrite \eqref{ceq1} and \eqref{ceq3} in vector notation to obtain constraints on the coefficients $c_n$ and $d_n$. Making use of the expression \eqref{Hfg}, one can show that eq. \eqref{ceq1} is fulfilled under the conditions
\bea
\pa^{\dot{\d} \a} G_{\a \bd,\, \g \gd, \, \d \dot{\d}} = 0~, \qquad 
\ve^{\a \d} G_{\a \bd,\, \g \gd, \, \d \dot{\d}} = 2 \ri \, \ve_{\dot{\d} \ad} \pa^{\ad \a} F_{\a \bd, \, \g \gd}~.
\eea
These differential constraints are equivalent to
\bsubeq 
\bea
&&\eta^{mp}\pa_k G_{mnp} - \pa^{m} (G_{mnk}+ G_{knm}) + \ri \e^{pqm}{}_{k} \pa_q G_{mnp} = 0~, \label{vectD1}\\
&&G_{mnp}- G_{pnm} = 2 \ri \, (\pa_p F_{mn} - \pa_m F_{pn})~, \quad \eta^{mp} G_{mnp} = 2\ri \, \pa^{m}F_{mn}~.\label{vectD2}
\eea
\esubeq
On the other hand, eq. \eqref{ceq3} implies that
\bsubeq 
\bea
\pa^{\dot{\b} \d} G_{\a \bd,\, \g \gd, \, \d \dot{\d}} = 0~, \qquad 
\ve^{\bd \dot{\d}} G_{\a \bd,\, \g \gd, \, \d \dot{\d}} = 2 \ri \, \ve_{{\d} \b} \pa^{\b \bd} F_{\a \bd, \, \g \gd}~,
\eea
\esubeq
which read
\bsubeq 
\bea
&&\eta^{mp}\pa_k G_{mnp} - \pa^{m} (G_{mnk}+ G_{knm}) - \ri \e^{pqm}{}_k \pa_q G_{mnp} = 0~, \label{vectQ1}\\
&&G_{mnp}- G_{pnm} = 2 \ri \, (\pa_p F_{mn} - \pa_m F_{pn})~, \quad \eta^{mp} G_{mnp} = 2\ri \, \pa^{m}F_{mn}~. \label{vectQ2}
\eea
\esubeq
Thus, eqs. \eqref{vectD2} and \eqref{vectQ2} are identical. It follows from eqs. \eqref{vectD1} and \eqref{vectQ1} that 
\bea
\e^{pqm}{}_k \pa_q G_{mnp} = 0~, \qquad \eta^{mp}\pa_k G_{mnp} - \pa^{m} (G_{mnk}+ G_{knm}) = 0~,
\eea
which lead to
\bea
d_5 = 0~, \qquad d_1 = -2 d_3~.
\eea
Upon imposing \eqref{vectD2}, we find that
\bea
c_1 &=& -\frac{1}{4} (d_2+ d_3)~, \qquad c_2 = \hf (d_3 + d_4)~. 
\eea
At this stage, we end up with three free coefficients, $d_2, d_3, d_4$ for our function $H_{mn}$:
\bea
H_{mn} (\bm{X}, \bm{\Q}, \bar{\bm{\Q}}) &=& -\frac{1}{4}(d_2+d_3) \frac{\eta_{mn}}{\bm{X}^2} + \hf (d_3 + d_4) \frac{\bm{X}_m \bm{X}_n}{(\bm{X}^2)^2} \non\\
&&+ \hf d_4 \frac{\ri}{(\bm{X}^2)^2} \bm{X}_m \bm{P}_n +  \hf d_3 \frac{\ri}{(\bm{X}^2)^2} \bm{X}_n \bm{P}_m \non\\
&&+ \hf d_2 \frac{\ri}{(\bm{X}^2)^2} \eta_{mn}(\bm{P} \cdot \bm{X} ) - d_3 \frac{\ri}{(\bm{X}^2)^3} (\bm{P} \cdot \bm{X}) \bm{X}_m \bm{X}_n~,
\eea
where we recalled that $\bm{P}_{ \a \ad} = -4 \bm{\Q}_{\a} \bm{\bar \Q}_{\ad}$.
We can also express our result in spinor notation:
\bea
H^{\bd \a}\,_{\g \gd} (\bm{X}, \bm{\Q}, \bar{\bm{\Q}}) &=& \frac{1}{2}(d_2+d_3) \frac{1}{\bm{X}^2} \d^{\a}_{\g} \,\d^{\bd}_{\gd} + \hf (d_3 + d_4) \frac{1}{(\bm{X}^2)^2}\bm{X}^{\bd \a} \bm{X}_{\g \gd}  \non\\
&&+ \hf d_4 \frac{\ri}{(\bm{X}^2)^2} \bm{X}^{\bd \a} \bm{P}_{\g \gd} +  \hf d_3 \frac{\ri}{(\bm{X}^2)^2} \bm{X}_{\g \gd} \bm{P}^{\bd \a} \non\\
&&- d_2 \frac{\ri}{(\bm{X}^2)^2} (\bm{P} \cdot \bm{X} ) \d^{\a}_{\g} \,\d^{\bd}_{\gd} - d_3 \frac{\ri}{(\bm{X}^2)^3} (\bm{P} \cdot \bm{X}) \bm{X}_{\g \gd} \bm{X}^{\bd \a}~. \label{Hspin}
\eea

It remains to impose conservation equations on the supercurrent $J_{\g \gd}$, eq.\,\eqref{ceq3rd}. 
Checking conservation laws on $z_3$ is non-trivial since there are no identities that  would allow differential
operators acting on the $z_3$ dependence to pass through the prefactor of \eqref{corr1}. 
To simplify our computation, let us first express our correlator \eqref{corr1} as
\bea \label{corr1-I}
\corr1 =
\frac{1}{k_1} I_{\a \ad}(x_{3 \bar{1}}) I_{\b \bd} (x_{2 \bar{3}}) H^{\bd \a}\,_{\g \gd} (\fn3)~,
\eea
where we have defined\footnote{The operator $I_{\a \ad} (x_{3 \bar{1}})$ was previously denoted by $(\hat{x}_{3 \bar{1}})_{\a \ad}$ in subsection 2.2.}
\be
\begin{aligned}
&I_{\a \ad} (x_{3 \bar{1}}) = \frac{(x_{3 \bar{1}})_{\a \ad}}{(x_{\bar{1} 3}{}^2)^{1/2}}~, \quad I_{\b \bd} (x_{2 \bar{3}}) = \frac{(x_{2 \bar{3}})_{\b \bd}}{(x_{\bar{3} 2}{}^2)^{1/2}}~,\\
&k_1 := (x_{\bar{1} 3}{}^2)^{3/2} (x_{\bar{3} 2}{}^2)^{3/2} x_{\bar{3} 1}{}^2 x_{\bar{2} 3}{}^2~.
\label{3.1}
\end{aligned}
\ee
On the other hand, the same correlator can be written as follows
\bea
&&\corr1 
= -\la {S}_{\b} (z_2) J_{\g \gd} (z_3) \bar{S}_{\ad} (z_1) \ra \non\\
&&= -\frac{1}{k_2} I_{\b \dot{\rho}}(x_{2 \bar{1}}) I_{\g \dot{\d}} (x_{3 \bar{1}}) I_{\d \dot{\g}}(x_{1 \bar{3}})\widetilde{H}_{\ad}^{\dot{\rho},\, \d \dot{\d}}(\bm{X}_1, \bm{\Q}_1, \bar{\bm{\Q}}_1)~,
\label{corr1-tr} \\
&&k_2 := (x_{\bar{1} 2}{}^2)^{3/2} (x_{\bar{1} 3}{}^2)^{3/2} (x_{\bar{3} 1}{}^2)^{3/2} x_{\bar{2} 1}{}^2~,
\non
\eea
for some function $\widetilde{H}_{\ad}^{\dot{\rho},\, \d \dot{\d}}(\bm{X}_1, \bm{\Q}_1, \bar{\bm{\Q}}_1) $. The second line in eq.~\eqref{corr1-tr}
has the same structure as the general expression~\eqref{3ptgen} except now we are treating $z_1$ as the ``third point". 
Since eqs.~\eqref{corr1-I} and~\eqref{corr1-tr} represent the same correlator the functions $H$ and $\widetilde{H}$ are related.
Knowing $H$ we can find $\widetilde{H}$ by comparing eqs.~\eqref{corr1-I} and~\eqref{corr1-tr}. 
For this let us introduce the operators inverse to~\eqref{3.1}:
\bea
\bar{I}^{\dot{\a} \a} (\tilde{x}_{\bar{1} 2}) = \frac{(x_{\bar{1} 2})^{\dot{\a} \a}}{(x_{\bar{1}2}{}^2)^{1/2}}~, \qquad 
\bar{I}^{\dot{\a} \b} (\tilde{x}_{\bar{1} 2}) I_{\b \dot{\b}} (x_{2 \bar{1}}) ={\d}_{\dot{\b}}\,^{\dot{\a}}~.
\label{3.2}
\eea
It follows from \eqref{corr1-I}, \eqref{3.1}, \eqref{corr1-tr} and~\eqref{3.2} that the transformed 
function $\widetilde{H}_{\ad}^{\dot{\rho}, \,\d \dot{\d}}$ 
can be directly computed from $H^{\bd \a}\,_{\g \gd} (\fn3)$ using the relation
\bea
&&\widetilde{H}_{\ad}^{\dot{\l}, \,\dot{\s} \s}(\bm{X}_1, \bm{\Q}_1, \bar{\bm{\Q}}_1)  
\non\\
&& =- \frac{k_2}{k_1} \Big[\bar{I}^{\dot{\l} \b} (\tilde{x}_{\bar{1} 2}) I_{\b \bd}(x_{2 \bar{3}}) \Big]\, 
\Big[ I_{\a \ad} (x_{3 \bar{1}}) \bar{I}^{\dot{\s} \g} (\tilde{x}_{\bar{1} 3}) \bar{I}^{\dot{\g} \s} (\tilde{x}_{\bar{3} 1}) \Big] H^{\bd \a}\,_{\g \gd} (\fn3)~. 
\label{3.3}
\eea
Using the identity 
\bea
\bar{I}^{\dot{\l} \b}(\tilde{x}_{\bar{1} 2}) I_{\b \dot{\mu}} (x_{2 \bar{3}}) \bar{I}^{\dot{\mu} \mu} (\tilde{x}_{\bar{3}1}) = \bar{I}^{\dot{\l} \mu} \big(\tilde{\bar {\bm{X}}}_1\big)~, \qquad 
\bar{I}^{\dot{\l} \mu} \big(\tilde{\bar {\bm{X}}}_1\big) = \frac{\bm{\bar X}_1^{\dot{\l} \mu}}{(\bm{\bar X}_1^2)^{1/2}}
\label{3.4}
\eea
we can also present eq.~\eqref{3.3} in the form
\bea
&&\widetilde{H}_{\ad}^{\dot{\l}, \,\dot{\s} \s}(\bm{X}_1, \bm{\Q}_1, \bar{\bm{\Q}}_1) 
\non\\
&&
 = -\frac{k_2}{k_1} \bar{I}^{\dot{\l} \mu}
(\tilde{\bar{\bm{X}}}_1)\Big[ I_{\a \ad}(x_{3 \bar{1}}) I_{\mu \bd} (x_{1 \bar{3}}) \bar{I}^{\dot{\s} \g} (\tilde{x}_{\bar{1} 3}) \bar{I}^{\dot{\g} \s} (\tilde{x}_{\bar{3} 1}) \Big]H^{\bd \a}\,_{\g \gd} (\fn3)~. \label{tildeH1}
\eea
To compute $\widetilde{H}_{\ad}^{\dot{\l}, \,\dot{\s} \s}$, we will use the following identities~\cite{OsbornN1}:
\bsubeq
\bea
&&
I_{\a \bd}(x_{3 \bar{1}}) I_{\b \ad}(x_{1 \bar{3}}) \bm{X}_3^{\ad \a} = 
-\frac{\bm{X}_1^{I}\,_{\b \bd}}{(\bm{X}_1^2 \bar{\bm X}_1^2 x_{\bar 1 3}{}^2 x_{\bar 3 1}{}^2)^{\hf}}~, \\
&&
{\bar I}^{\ad \b}(x_{\bar{3} 1}) {\bar I}^{\bd \a}(x_{\bar{1} 3}) \bm{X}_{3\, \a \ad} = -\frac{\bm{X}_1^{I\,\bd \b}} {(\bm{X}_1^2 \bar{\bm X}_1^2 x_{\bar 1 3}{}^2 x_{\bar 3 1}{}^2)^{\hf}}~, \\
&&
I_{\a \ad}(x_{1 \bar{3}}) \bar{\bm \Q}_3^{\ad} = \bigg( \frac{x_{\bar{3} 1}{}^2}{\bar{\bm X}_1^2} \bigg)^\hf \frac{\bm{\bar \Q}_{1 \a}^I}{x_{\bar{1} 3}{}^2}~,
\qquad 
\bm{\Q}_3^{\a} I_{\a \ad}(x_{3 \bar{1}}) = \bigg( \frac{x_{\bar{1} 3}{}^2}{\bm{X}_1^2} \bigg)^\hf \frac{\bm{\Q}_{1 \ad}^I}{x_{\bar{3} 1}{}^2}~.
\eea
Here we have defined
\bea
&&
\bm{X}_1^{I}\,_{\a \ad} := -\Big(\frac{\bm{X}_1^2}{\bar{\bm X}_1^2}\Big)^{\hf} \bar{\bm X}_{1\, \a \ad}~, \\
&&
\bm{\Q}_{1 \ad}^{I} := \bm{\Q}_1^{\a} I_{\a \ad}(-\bm{X}_1)  =- \frac{\bm{\Q}_1^{\a} \bm{X}_{1 \a \ad}}{(\bm{X}_1^2)^\hf}~,  \\
&&
\bar{\bm \Q}_{1 \a}^{I} := I_{\a \ad}(\bar{\bm X}_1) \bar{\bm \Q}_1^{\ad} = \frac{\bar{\bm X}_{1 \a \ad}}{(\bar{\bm X}_1^2)^\hf} \bar{\bm \Q}_1^{\ad}~.
\eea
\esubeq
After some calculations using the above identities we
find that $\widetilde{H}_{\ad}^{\dot{\rho}, \d \dot{\d}}(\bm{X}_1, \bm{\Q}_1, \bar{\bm{\Q}}_1)$ takes the form
\be
\begin{aligned}
&\widetilde{H}_{\ad}^{\dot{\rho}, \,\dot{\d} \d}(\bm{X}_1, \bm{\Q}_1, \bar{\bm{\Q}}_1) = -\frac{1}{2}(d_2+d_3) \frac{1}{\bm{X}_1^2 \bar{\bm{X}}_1^2} \bar{\bm{X}_1}^{\dot{\rho} \d} \,\d^{\dot{\d}}_{\ad} + \frac{1}{2}(d_3 + d_4) \frac{1}{\bm{X}_1^2 \bar{\bm{X}}_1^2} \bar{\bm{X}}_1^{\dot{\d} \d} \,\d^{\dot{\rho}}_{\ad} \\
&\qquad - \frac{\ri}{2} d_4 \frac{1}{(\bm{X}_1^2)^2 \bar{\bm{X}}_1^2} \bar{\bm{X}}_1^{\dot{\b} \d} \bm{P}_{1 \b \bd} \bm{X}_1^{\dot{\d} \b} \,\d^{\dot{\rho}}_{\ad} - \frac{\ri}{2} d_3 \frac{1}{(\bm{X}_1^2)^2 \bar{\bm{X}}_1^2} \bar{\bm{X}}_1^{\dot{\d} \d} \bm{X}_{1 \b \ad} \bm{P}_1^{\dot{\rho} \b} \\
&\qquad +\ri\, d_2 \frac{1}{(\bm{X}_1^2)^2 \bar{\bm{X}}_1^2} \bar{\bm{X}}_1^{\dot{\rho} \d} (\bm{P}_1 \cdot \bm{X}_1 )\,\d^{\dot{\d}}_{\ad} -\ri\, d_3 \frac{1}{(\bm{X}_1^2)^2 \bar{\bm{X}}_1^2} \bar{\bm{X}}_1^{\dot{\d} \d} (\bm{P}_1 \cdot \bm{X}_1 )\,\d^{\dot{\rho}}_{\ad}~.~~~ \label{326}
\end{aligned}
\ee
Next, relabelling the superspace points 
$z_2\rightarrow z_1, \,z_3 \rightarrow z_2, \,z_1 \rightarrow z_3$, along with eqs.\,\eqref{Z13}, allows us to rewrite \eqref{corr1-tr} 
in the equivalent form
\bea
&&
\la {S}_{\b} (z_1) J_{\g \gd} (z_2) \bar{S}_{\ad} (z_3)\ra = \frac{1}{k_3} I_{\b \dot{\rho}}(x_{1 \bar{3}}) I_{\g \dot{\d}} (x_{2 \bar{3}}) I_{\d \dot{\g}}(x_{3 \bar{2}})\widetilde{H}_{\ad}^{\dot{\rho}, \d \dot{\d}}(\bm{X}_3, \bm{\Q}_3, \bm{\bar{\Q}}_3)~, \label{corr1-tr2} \\
&&
k_3 := (x_{\bar{3} 1}{}^2)^{3/2} (x_{\bar{3} 2}{}^2)^{3/2} (x_{\bar{2} 3}{}^2)^{3/2} x_{\bar{1} 3}{}^2~.
\non
\eea
Now we can apply the conservation condition of $J_{\g \gd}$ using eqs.~\eqref{Cderivs}. 
We get
\bea
\cQ_{\d}\widetilde{H}_{\ad}^{\dot{\rho}, \d \dot{\d}}(\bm{X}_3, \bm{\Q}_3, \bm{\bar{\Q}}_3)= 0~, \qquad \bar \cQ_{\dot{\d}}\widetilde{H}_{\ad}^{\dot{\rho}, \d \dot{\d}}(\bm{X}_3, \bm{\Q}_3, \bm{\bar{\Q}}_3)= 0~. \label{325}
\eea
After some calculations, one may verify that \eqref{326} satisfies \eqref{325} for an arbitrary choice of $d_2, d_3, d_4$. 

Finally, we note that our three-point function $\corr1$ satisfies the reality condition 
\be 
\la \bar S_{\ad}(z_1) S_{\b}(z_2) J_{\g \gd}(z_3) \ra = \la \bar S_{\bd}(z_2) S_{\a}(z_1) J_{\g \gd}(z_3) \ra^*~,
\ee
where $*$ means complex conjugation. 
It implies the following condition on the function $\bar{H}^{\ad \b}\,_{\g \gd}$:
\bea
\bar{H}^{\ad \b}\,_{\g \gd} (\fn3) = H^{\ad \b}\,_{\g \gd} (-\bm{\bar X}_3, -\bm{\Q}_3, -\bar{\bm \Q}_3)~,
\eea
where $\bar{H}^{\ad \b}\,_{\g \gd} (\fn3)$ is the complex conjugate of the expression \eqref{Hspin}.
This condition gives the reality constraint on the coefficients:
\bea
\bar{d}_2 = d_2 ~, \qquad \bar{d}_3 = d_3~, \qquad \bar{d}_4 = d_4~.
\eea
Therefore, our final expression for the three-point function $\corr1$ is determined up to three independent, real coefficients:
\bsubeq
\label{corr1fin}
\bea
\corr1 &=& \frac{I_{\a \ad}(x_{3 \bar 1}) I_{\b \bd}(x_{2 \bar 3})}{(x_{\bar 1 3}{}^2)^{3/2} (x_{\bar 3 2}{}^2)^{3/2} x_{\bar 3 1}{}^2 x_{\bar 2 3}{}^2} 
H^{\bd \a}\,_{\g \gd} (\fn3)~,
\eea
with
\bea 
&&H^{\bd \a}\,_{\g \gd} (\bm{X}, \bm{\Q}, \bm{\bar{\Q}}) = \frac{1}{2}(d_2+d_3) \frac{1}{\bm{X}^2} \d^{\a}_{\g} \,\d^{\bd}_{\gd} + \hf (d_3 + d_4) \frac{1}{(\bm{X}^2)^2}\bm{X}^{\bd \a} \bm{X}_{\g \gd}  \non\\
&&\qquad \quad + \hf d_4 \frac{\ri}{(\bm{X}^2)^2} \bm{X}^{\bd \a} \bm{P}_{\g \gd} +  \hf d_3 \frac{\ri}{(\bm{X}^2)^2}  \bm{P}^{\bd \a} \bm{X}_{\g \gd} \non\\
&&\qquad \quad - d_2 \frac{\ri}{(\bm{X}^2)^2} (\bm{P} \cdot \bm{X} ) \d^{\a}_{\g} \,\d^{\bd}_{\gd} - d_3 \frac{\ri}{(\bm{X}^2)^3} (\bm{P} \cdot \bm{X})  \bm{X}^{\bd \a} \bm{X}_{\g \gd}~.
\eea
In vector notation, this is equivalent to
\bea
&&H_{mn} (\bm{X}, \bm{\Q}, \bar{\bm{\Q}}) = \frac{1}{4} (\s_m)_{\a \bd} (\tilde{\s}_n)^{\gd \g}H^{\bd \a}\,_{\g \gd} (\bm{X}, \bm{\Q}, \bm{\bar{\Q}})\non\\
&&\quad \quad =  -\frac{1}{4}(d_2+d_3) \frac{\eta_{mn}}{\bm{X}^2} + \hf (d_3 + d_4) \frac{\bm{X}_m \bm{X}_n}{(\bm{X}^2)^2} \non\\
&&\quad \quad + \hf d_4 \frac{\ri}{(\bm{X}^2)^2} \bm{X}_m \bm{P}_n +  \hf d_3 \frac{\ri}{(\bm{X}^2)^2} \bm{X}_n \bm{P}_m \non\\
&&\quad \quad + \hf d_2 \frac{\ri}{(\bm{X}^2)^2} \eta_{mn}(\bm{P} \cdot \bm{X} ) - d_3 \frac{\ri}{(\bm{X}^2)^3} (\bm{P} \cdot \bm{X}) \bm{X}_m \bm{X}_n~,
\eea
\esubeq
Later we will see that eq.~\eqref{corr1fin} cannot be obtained from three-point function of the $\cN=2$ supercurrent unless
the coefficients $d_2, d_3, d_4$ satisfy an additional constraint $d_2+ d_3+ d_4=0$. However, $\cN=1$ superconformal symmetry alone
does not impose this constraint.


\subsection{Correlator $\la S_{\a}(z_1) J_{\b \bd}(z_2) S_{\g}(z_3) \ra$} \label{ss32}


In the remaining part of the section we will consider two other correlators involving the supercurrent and spinor currents. 
They both have non-compensating $R$-symmetry charge and are expected to vanish. We will check that it is indeed the case. 

According to \eqref{3ptgen}, we write the general ansatz
\bea
\la S_{\a}(z_1) J_{\b \bd}(z_2) S_{\g}(z_3) \ra = \frac{(x_{1 \bar{3}})_{\a \ad} (x_{2 \bar{3}})_{\b \dot{\d}} (x_{3 \bar{2}})_{\d \dot{\b}} }{x_{\bar{1} 3}{}^2 (x_{\bar{3} 1}{}^2)^2 (x_{\bar{2} 3}{}^2)^2 (x_{\bar{3} 2}{}^2)^2 } \, H^{\ad, \, \dot{\d} \d}\,_{\g}(\fn3)~.
\eea
As a result of imposing the conservation equations \eqref{c1} and \eqref{c2}, the tensor $H^{\ad, \dot{\d} \d}\,_{\g}$ must satisfy the following constraints
\be
\begin{aligned}
&\bar{\cQ}^{\dot{\d}}H_{\ad,\, \d \dot{\d}, \,\g} = 0~, \qquad {\cQ}^{{\d}}H_{\ad,\, \d \dot{\d}, \,\g} = 0~,\\
&\bar{\cD}^{\dot{\a}}H_{\ad,\, \d \dot{\d}, \,\g} = 0~, \qquad {\cD^2}H_{\ad,\, \d \dot{\d}, \,\g} = 0~.
\end{aligned}
\ee
It obeys the homogeneity property
\bea \label{hom2}
H^{\ad,\, \dot{\d} \d}\,_{\g}(\l \bar{\l} \bm{X}, \l \bm{\Q}, \bar{\l} \bar{\bm \Q} ) = \l^{-3} \bar \l^{-3} H^{\ad, \,\dot{\d} \d}\,_{\g} (\bm{X}, \bm{\Q}, \bar{\bm \Q})~.
\eea
We also need to make sure that the correlator has the right symmetry property under the replacement of superspace points $z_1 \leftrightarrow z_3$,
\bea
\la S_{\a}(z_1) J_{\b \bd}(z_2) S_{\g}(z_3) \ra = - \la S_{\g}(z_3) J_{\b \bd}(z_2) S_{\a}(z_1) \ra~,
\eea
since the superfield $S_{\a}$ is Grassmann odd. 

As in the previous subsection, it is useful to switch to vector notation,
\bea
H_{\ad,\, \d \dot{\d}, \g} = (\s^m)_{\d \dot{\d}} (\s^n)_{\g \ad} H_{mn}~.
\eea
Since $H_{mn} (\fxq)$ is Grassmann even, the most general form we can write is
\be
\begin{aligned}
H_{mn} (\fxq) &= A_{mn} (\bm{X}) + \bm{\Q}^2 B_{mn} (\bm{X}) + \bar{\bm \Q}^2 C_{mn} (\bm{X}) + \bm{\Q}^2 \bar{\bm \Q}^2 D_{mn}(\bm{X}) \\
&+ \bm{\Q}^{\a} \bar{\bm{\Q}}^{\bd} (\s^p)_{\a \bd} F_{mnp} (\bm{X})~.
\end{aligned}
\ee
The requirements ${\cQ}^{{\d}}H_{\ad,\, \d \dot{\d}, \,\g} = 0$ and $\bar{\cD}^{\dot{\a}}H_{\ad,\, \d \dot{\d}, \,\g}=0$ lead to the algebraic constraints
\bsubeq
\bea
&&B_{mn} = 0~, \qquad C_{mn}=0~, \qquad D_{mn} = 0~,\\
&& \eta^{mp}F_{mnp} = \eta^{np}F_{mnp} = 0~, \qquad F_{mnp} = F_{mpn} = F_{pnm}~, \label{cor2-alg2}
\eea
\esubeq
and hence, \eqref{cor2-alg2} implies that $F_{mnp} (\bm{X})$ is completely symmetric and traceless.
On the other hand, imposing $\bar{\cQ}^{\dot{\d}}H_{\ad,\, \d \dot{\d}, \,\g} = 0$ and $\cD^2 H_{\ad,\, \d \dot{\d}, \,\g} = 0$ gives us several differential constraints:
\bsubeq
\bea
&&\pa^m F_{mnp} = 0~, \label{cor2-dif1}\\
&&\pa^m A_{mn} = \pa_{[p}A_{m]n} = 0~, \qquad \Box A_{mn}=0~,
\eea
\esubeq
where $\Box := \pa^m \pa_m$.

Let us solve for $F_{mnp}(\bm{X})$. Being completely symmetric in $m,n,p$, the general solution for $F_{mnp}(\bm{X})$ compatible with the homogeneity property \eqref{hom2} reads
\bea
F_{mnp} = \frac{f_1}{(\bm{X}^2)^{7/2}} \bm{X}_m \bm{X}_n \bm{X}_p +  \frac{f_2}{(\bm{X}^2)^{5/2}} \Big(\eta_{mn} \bm{X}_p + \eta_{np} \bm{X}_m + \eta_{mp} \bm{X}_n \Big)
\eea
Imposing \eqref{cor2-alg2} and \eqref{cor2-dif1} leads to
\bea
f_1 = f_2 = 0 ~.
\eea
Next, we turn to $A_{mn}(\bm{X})$, for which we have the general expression
\bea
A_{mn} = a_1 \frac{\eta_{mn}}{(\bm{X}^2)^{3/2}} + a_2 \frac{\bm{X}_m \bm{X}_n}{(\bm{X}^2)^{5/2}}~.
\eea
It is easy to check that the conditions $\pa^m A_{mn} =0$ and $\pa_{[p}A_{m]n} = 0$ already give 
\bea
a_1 = a_2 = 0~.
\eea
Therefore, we conclude that
\bea
\la S_{\a}(z_1) J_{\b \bd}(z_2) S_{\g}(z_3) \ra = 0~.
\eea


\subsection{Correlator $\la S_{\a}(z_1) J_{\b \bd}(z_2) J_{\g \gd}(z_3) \ra$}


According to eq.~\eqref{3ptgen}, the general expression of this three-point function is as follows 
\bea
\la S_{\a}(z_1) J_{\b \bd}(z_2) J_{\g \gd}(z_3) \ra = \frac{(x_{1 \bar{3}})_{\a \ad} (x_{2 \bar{3}})_{\b \dot{\d}} (x_{3 \bar{2}})_{\d \dot{\b}} }{x_{\bar{1} 3}{}^2 (x_{\bar{3} 1}{}^2)^2 (x_{\bar{2} 3}{}^2)^2 (x_{\bar{3} 2}{}^2)^2 } \, H^{\ad, \, \dot{\d} \d}\,_{\g \gd}(\fn3)~.
\eea
As a result of imposing the conservation equations \eqref{c1} and \eqref{c2}, the tensor $H^{\ad,\, \dot{\d} \d}\,_{\g \gd}$ must satisfy the following constraints
\be
\begin{aligned}
&\bar{\cQ}^{\dot{\d}}H_{\ad,\, \d \dot{\d}, \,\g \gd} = 0~, \qquad {\cQ}^{{\d}}H_{\ad,\, \d \dot{\d}, \,\g \gd} = 0~,\\
&\bar{\cD}^{\dot{\a}}H_{\ad,\, \d \dot{\d}, \,\g \gd} = 0~, \qquad {\cD^2}H_{\ad,\, \d \dot{\d}, \,\g \gd} = 0~.
\end{aligned}
\ee
It also obeys the homogeneity property
\bea \label{hom3}
H^{\ad,\, \dot{\d} \d}\,_{\g \gd}(\l \bar{\l} \bm{X}, \l \bm{\Q}, \bar{\l} \bar{\bm \Q} ) = \l^{-8/3} \bar \l^{-7/3} H^{\ad, \,\dot{\d} \d}\,_{\g \gd} (\fxq)~.
\eea
The correlator also satisfies the 
symmetry under $z_2 \leftrightarrow z_3$:
\bea
\la S_{\a}(z_1) J_{\b \bd}(z_2) J_{\g \gd}(z_3) \ra = \la S_{\a}(z_1) J_{\g \gd}(z_3) J_{\b \bd}(z_2)  \ra~.
\eea

Let us now write
\bea
H_{\ad,\, \d \dot{\d}, \g \gd} = (\s^m)_{\d \dot{\d}} (\s^n)_{\g \gd} H_{mn,\, \ad}~.
\eea
Since $H_{mn,\, \ad} (\fxq)$ is Grassmann odd, the most general form we can have is
\be
\begin{aligned}
H_{mn,\, \ad} (\fxq) &= \bm{\Q}^{\rho} (\s^k)_{\rho \ad} A_{kmn} (\bm{X}) + \bar{\bm \Q}^2 \bm{\Q}^{\rho} (\s^k)_{\rho \ad} B_{kmn} (\bm{X})\\ 
&+\bar{\bm \Q}^{\dot{\rho}} C_{mn, \,\dot{\rho}, \ad} (\bm{X}) + \bm{\Q}^2 \bar{\bm \Q}^{\dot{\rho}} D_{mn, \,\dot{\rho}, \ad} (\bm{X})~.
\end{aligned}
\ee
Imposing ${\cQ}^{{\d}}H_{\ad,\, \d \dot{\d}, \,\g \gd} = 0$ and $\bar{\cD}^{\dot{\a}}H_{\ad,\, \d \dot{\d}, \,\g \gd}=0$ leads to the requirements
\bsubeq
\bea
&&B_{kmn} = 0~, \qquad D_{mn,\, \dot{\rho}, \ad}=0~,\\
&& C_{mn,\, \dot{\rho}, \ad} = \hf \ve_{\dot{\r} \dot{\l}} \big( \tilde{\s}^k \s^p - \tilde{\s}^p \s^k \big)^{\dot{\l}}\,_{\ad} \,\widetilde{C}_{mn\,[kp]}~, \\
&& \eta^{km} A_{kmn} = 0~, \qquad A_{[km]\,n} = 0~. \label{corr3-alg3}
\eea
\esubeq
The condition $\bar{\cQ}^{\dot{\d}}H_{\ad,\, \d \dot{\d}, \,\g \gd} = 0$ is equivalent to
\bsubeq
\bea
&& \big( \s^m \tilde{\s}^k \s^p \big) \tilde{C}_{mn\,[kp]} = 0~, \quad \pa^m \widetilde{C}_{mn\,[kp]} = 0~, \quad \pa_{[r}\widetilde{C}_{m]n\,[kp]} = 0~,\\
&& \pa^{k} A_{kmn}=0~, \qquad \pa_{[r} A_{k]mn}=0~.
\eea
\esubeq
It can be shown that ${\cD^2}H_{\ad,\, \d \dot{\d}, \,\g \gd} = 0$ does not give further constraints.

Let us try to construct the explicit solution for $A_{kmn}(\bm{X})$. Since it has dimension $-3$, we can have a general expression
\be
\begin{aligned}
A_{kmn} = \frac{a_1}{(\bm{X}^2)^{3}} \bm{X}_k \bm{X}_m \bm{X}_n  + &\frac{1}{(\bm{X}^2)^{2}} \Big( a_2 \,\eta_{km} \bm{X}_n + a_3 \,\eta_{kn} \bm{X}_m + a_4 \,\eta_{mn} \bm{X}_k \\
&+ a_5 \,\e_{kmnp}\bm{X}^p \Big)
\end{aligned}
\ee
Eq. \eqref{corr3-alg3} relates the coefficients in the following way
\bea
a_5 =0~, \qquad  a_3 = a_4 = -\hf(a_1 + 4 a_2)~.
\eea
Demanding that $\pa^{k} A_{kmn}=0$ leads to $a_2 = -\hf a_1$, while $\pa_{[r} A_{k]mn}=0$ gives $a_1 = 0$. As a result,
\bea
A_{kmn} (\bm{X}) = 0 ~.
\eea
Next, we solve for $\widetilde{C}_{mn\, [kp]}(\bm{X})$, for which we have the general expression
\be
\begin{aligned}
\widetilde{C}_{mn\, [kp]} &= \frac{c_1}{(\bm{X}^2)^{3/2}} \,\e_{mnkp}+  \frac{c_2}{(\bm{X}^2)^{3/2}} \big( \eta_{mk} \eta_{np} - \eta_{mp} \eta_{nk} \big) \\
&+ \frac{c_3}{(\bm{X}^2)^{5/2}} \big( \eta_{nk} \bm{X}_p \bm{X}_m - \eta_{np} \bm{X}_k \bm{X}_m \big) + \frac{c_4}{(\bm{X}^2)^{5/2}} \big( \eta_{mk} \bm{X}_p \bm{X}_n - \eta_{mp} \bm{X}_k \bm{X}_n \big)~. \label{corr2-C}
\end{aligned}
\ee
It can be shown that \eqref{corr2-C} solves $\big( \s^m \tilde{\s}^k \s^p \big) \tilde{C}_{mn\,[kp]} = 0$ for
\bea
c_1 = \ri (c_2 - \frac{c_3}{3})~, \qquad  c_4 = -\frac{2}{3} c_3~. \label{cc1}
\eea
In deriving the above, we have made use of the identity
\bea
\s^m \tilde{\s}^k \s^p = \eta^{mp}\s^k - \eta^{kp}\s^m - \eta^{mk}\s^p  + \ri \e^{mkpq} \s_q~.
\eea
On the other hand, the constraint $\pa^m \widetilde{C}_{mn\,[kp]} = 0$ requires
\bea
c_1 = 0~, \qquad c_4 = -3 c_2~. \label{cc2}
\eea
Taking into account \eqref{cc1}, we immediately obtain
\bea
\widetilde{C}_{mn\,[kp]}(\bm{X}) = 0~.
\eea
Therefore, we conclude that
\bea
\la S_{\a}(z_1) J_{\b \bd}(z_2) J_{\g \gd}(z_3) \ra = 0~.
\eea


\section{Correlation functions of spinor current with the flavour current multiplets}  \label{Section4}


The $\cN=1$ flavour current multiplet is a primary real superfield $L^{\bar a} = \bar{L}^{\bar a}$, 
where $\bar a$ is the index of a flavour symmetry group, 
subject to the conservation equation 
\bea
D^2 L^{\bar a} = \bar{D}^2 L^{\bar a} = 0~.
\label{flavour-ce}
\eea
Its superconformal transformation law is
\be
\d L = -\xi L -2 (\s + \bar{\s}) L~.
\ee
It then follows that it has weights $(q, \bar q)= (1, 1)$ and dimension 2. 

In this section, we consider three-point functions of spinor current multiplets $S_{\a}$, ${\bar S}_{\dot{\a}}$ with a $\rm U(1)$ flavour current $L$. Generalisations to the non-Abelian case are also straightforward. 


\subsection{Correlator $\la \bar S_{\ad}(z_1) S_{\b}(z_2) L (z_3) \ra$}


As in the previous section, we start with the general form of the three-point function 
\be
\begin{aligned}
\la \bar S_{\ad}(z_1) S_{\b}(z_2) L(z_3) \ra 
&= \frac{(x_{3 \bar{1}})_{\a \ad} (x_{2 \bar{3}})_{\b \bd}}{(x_{\bar{1} 3}{}^2)^{2} x_{\bar{3}1}{}^2 x_{\bar{2} 3}{}^2(x_{\bar{3} 2}{}^2)^{2}} H^{\bd \a}(\fn3)~,
\end{aligned}
\label{JJL}
\ee
where $H^{\bd \a}$ has the following homogeneity property
\bea
H^{\bd \a}(\l \bar{\l} \bm{X}, \l \bm{\Q}, \bar{\l} \bar{\bm \Q} ) = \l^{-3} \bar \l^{-3} H^{\bd \a} (\fxq)~.
\label{scale}
\eea
By taking the complex conjugate of \eqref{JJL}, we have the reality condition
\be
\bar{H}^{\ad \b}(\fn3) = H^{\ad \b}(-\bm {\bar X}_3, -{\bm \Q}_3, -\bar {\bm \Q}_3)~.
\label{re}
\ee
As a consequence of the spinor current conservation law \eqref{c1}(and its conjugate), the tensor $H_{\a \bd}$ must satisfy the differential constraints
\be
\begin{aligned}
&\cD^{\a}H_{\a \bd}=0~, \qquad \bar \cD^2 H_{\a \bd}=0~, \\
& \bar{\cQ}^{\bd}H_{\a \bd}=0~, \qquad \cQ^2 H_{\a \bd}=0~.
\end{aligned}
\label{deq}
\ee
The conservation equations for the flavour current,
\bea
D^2_{(3)}\la \bar S_{\ad}(z_1) S_{\b}(z_2) L(z_3) \ra = 0~, \qquad \bar D^2_{(3)}\la \bar S_{\ad}(z_1) S_{\b}(z_2) L(z_3) \ra = 0
\label{4.0}
\eea
will be imposed at the end, by making use of a similar trick as in subsection \ref{Subsec3.1}.

Since $H_{\a \bd}$ is Grassmann even we have the following expansion 
\bea
H_{\a \bd}(\fxq) &= & 
F_{\a \bd} (\bm{X}) + A_{\a \bd} (\bm{X}) {\bm \Q}^2 + B_{\a \bd} (\bm{X}) \bar{\bm \Q}^2
+ C_{\a \bd} (\bm{X})  {\bm \Q}^2 \bar{\bm \Q}^2 \non\\
&+& G_{\a \bd, \d \dot{\d}} (\bm{X}){\bm \Q}^{\d}  \bar{\bm \Q}^{\dot{\d}}\,. 
\label{4.1}
\eea
From the conditions $\bar \cD^2 H_{\a \bd}=0, \ \cQ^2 H_{\a \bd}=0$ we quickly obtain that 
\be 
A_{\a \bd} (\bm{X}) = B_{\a \bd} (\bm{X})= C_{\a \bd} (\bm{X})=0
\label{4.2}
\ee
which leaves us with 
\be 
H_{\a \bd}(\bm{X}, {\bm \Q}, \bar {\bm \Q}) =F_{\a \bd} (\bm{X}) + G_{\a \bd, \d \dot{\d}} (\bm{X}){\bm \Q}^{\d}  \bar{\bm \Q}^{\dot{\d}}\,.
\label{4.3}
\ee
Converting to vector notation 
\be 
F_{\a \bd}= (\s^m)_{\a \bd} F_m\,, \qquad G_{\a \bd, \d \dot{\d}}=
(\s^m)_{\a \bd}(\s^n)_{\d \dot{\d}} G_{mn}~,
\label{4.4}
\ee
we see that there are only few possible options for $F_m$ and $G_{mn}$:
\be 
F_m= c_1\frac{\bm{X}_m}{(\bm{X}^2)^2}\,, \qquad G_{mn}= d_1 \frac{\eta_{m n}}{(\bm{X}^2)^2} + d_2 \frac{\bm{X}_m\bm{X}_n}{(\bm{X}^2)^3}\,, 
\label{4.5}
\ee
where $c_1, d_1, d_2$ are arbitrary free coefficients. Substituting eqs.~\eqref{4.3}, \eqref{4.4}, \eqref{4.5} into eq.~\eqref{deq}
we find that $d_2=- 4 d_1$. Using the reality property~\eqref{re} we find that $c_1$ is imaginary and $d_1$ is real. 
Hence, we have the following solution
\be
H_{\a \bd} = \ri c_1 \frac{\bm X_{\a \bd}}{(\bm X^2)^2}
+ \frac{d_1}{2 (\bm X^2)^3} \Big( \bm X^2 \bm P_{\a \bd} - 4 (\bm P \cdot \bm X) \bm X_{\a \bd} \Big)~,
\label{H-JJL}
\ee
where we redefined $c_1 \to \ri c_1$ so that 
$c_1$ and $d_1$ are now real, free coefficients.

We still must check the flavour current conservation equations~\eqref{4.0}. First, we will express our correlator \eqref{JJL} as
\bea
&&
\la \bar S_{\ad}(z_1) S_{\b}(z_2) L(z_3) \ra = \frac{1}{k_1} I_{\a \ad}(x_{3 \bar{1}}) I_{\b \bd} (x_{2 \bar{3}}) H^{\bd \a} (\fn3)~,
\label{4.6}
\\
&&
k_1 := (x_{\bar{1} 3}{}^2)^{3/2} x_{\bar{3} 1}{}^2 x_{\bar{2} 3}{}^2 (x_{\bar{3} 2}{}^2)^{3/2}~.
\non
\eea
By rearranging the operators in the three-point function, we may write
\bea
&&
\la \bar S_{\ad}(z_1) S_{\b}(z_2) L(z_3) \ra = - \la S_{\b}(z_2) L(z_3) \bar S_{\ad}(z_1) \ra 
= -\frac{1}{k_2} I_{\b \gd}(x_{2 \bar{1}}) \widetilde{H}^{\gd}{}_{\ad}(\bm X_1, \bm \Q_1, \bar {\bm \Q}_1)~, 
\non \\
&&
k_2 := (x_{\bar{1} 2}{}^2)^{3/2} x_{\bar{2} 1}{}^2 x_{\bar{1} 3}{}^2 x_{\bar{3} 1}{}^2~.
\label{JJL-re}
\eea
Comparing eqs.~\eqref{4.6} and~\eqref{JJL-re} we find  the relation between $H$ and $\widetilde{H}$:
\be
\widetilde{H}^{\gd}{}_{\ad}(\bm X_1, \bm \Q_1, \bar {\bm \Q}_1) = -\frac{k_2}{k_1} \bar{I}^{\gd \d}
(\tilde{\bar{\bm X}}_1)\Big[ I_{\d \bd} (x_{1 \bar{3}}) I_{\a \ad}(x_{3 \bar{1}}) H^{\bd \a} (\fn3) \Big]~,
\label{tildeHH0}
\ee
where the identity~\eqref{3.4} has also been used. Performing a similar calculation as in the previous section we find 
\be
\widetilde{H}^{\gd}{}_{\ad}(\bm X_1, \bm \Q_1, \bm {\bar \Q}_1) 
= \frac{\ri c_1}{\bm X_1^2} \d^{\gd}{}_{\ad} - \frac{d_1}{2 (\bm X_1^2)^2} \Big( \bm P_1^{\gd \g} 
\bm X_{1\g \ad} + 4 \d^{\gd}{}_{\ad} (\bm P_1 \cdot \bm X_1)\Big)~. 
\label{tildeHH}
\ee
Upon relabelling superspace points $z_2\rightarrow z_1, \,z_3 \rightarrow z_2, \,z_1 \rightarrow z_3$, eq. \eqref{JJL-re} turns into
\bea
&&
\la S_{\b}(z_1) L(z_2) \bar S_{\ad}(z_3) \ra
= \frac{1}{k_3} I_{\b \gd}(x_{1 \bar{3}}) \widetilde{H}^{\gd}{}_{\ad}(\bm X_3, \bm \Q_3, \bm {\bar \Q}_3)~,
\label{4.7} \\
&&
k_3 := (x_{\bar{3} 1}{}^2)^{3/2} x_{\bar{3} 2}{}^2 x_{\bar{2} 3}{}^2 x_{\bar{1} 3}{}^2~.
\non
\eea
Now the conservation conditions~\eqref{4.0} become
\bea
\cQ^2 \widetilde{H}^{\gd}{}_{\ad}(\bm X_3, \bm \Q_3, \bar {\bm \Q}_3)= 0~, 
\qquad \bar \cQ^2 \widetilde{H}^{\gd}{}_{\ad}(\bm X_3, \bm \Q_3, \bar {\bm \Q}_3)=0~.
\label{4.8}
\eea
By explicit calculations one can show that $\widetilde{H}$ in eq.~\eqref{tildeHH} satisfies~\eqref{4.8} 
for arbitrary $c_1$ and $d_1$. 

Thus, in $\cN=1$ superconformal theory, the three-point function involving spinor current multiplets and a flavour current superfield has two linearly independent functional structures with real coefficients $c_1$ and $d_1$:
\bea
\la \bar S_{\ad}(z_1) S_{\b}(z_2) L(z_3) \ra 
= \frac{(x_{3 \bar{1}})_{\a \ad} (x_{2 \bar{3}})_{\b \bd}}{(x_{\bar{1} 3}{}^2)^{2} x_{\bar{3}1}{}^2 x_{\bar{2} 3}{}^2(x_{\bar{3} 2}{}^2)^{2}} H^{\bd \a}(\fn3)~,
\label{JJL-fin}
\eea
where
\be
H_{\a \bd} = \ri c_1 \frac{\bm X_{\a \bd}}{(\bm X^2)^2}
+ \frac{d_1}{2 (\bm X^2)^3} \Big( \bm X^2 \bm P_{\a \bd} - 4 (\bm P \cdot \bm X) \bm X_{\a \bd} \Big)~.
\label{tensor-JJL}
\ee
%
In section \ref{Section6}, we will study how this correlator is related to ${\cN}=2$ superconformal symmetry. 


\subsection{Correlator $\la L(z_1) L(z_2) S_{\a}(z_3) \ra$}


This correlator carries a non-vanishing $R$-symmetry charge and we will show that it vanishes. 
Here we demonstrate that by imposing the conservation laws for $L(z_1)$ and $L(z_2)$, it is sufficient to see that 
$\la L(z_1) L(z_2) S_{\a}(z_3) \ra = 0$. As usual, we start with the general form
\bea
\la L(z_1) L(z_2) S_{\a}(z_3) \ra = \frac{1}{x_{\bar{1} 3}{}^2 x_{\bar{3} 1}{}^2 x_{\bar{2} 3}{}^2 x_{\bar{3} 2}{}^2}\, H_{\a} (\fn3)~,
\eea
where the tensor $H_{\a} (\fn3)$ obeys 
\bea
H_{\a}(\l \bar{\l} \bm{X}, \l \bm{\Q}, \bar{\l} \bar{\bm \Q} ) = \l^{-5/3} \bar \l^{-4/3} H_{\a}(\fxq)~, \label{hom4}
\eea
and so its dimension is $-3/2$. 
The conservation equations for $J(z_1)$ and $J(z_2)$ result in 
\be
\begin{aligned}
&\bar{\cD}^{2}H_{\a} = 0~, \qquad \cD^2 H_{\a} = 0~,\\
& \bar{\cQ}^2 H_{\a} = 0~, \qquad \cQ^2 H_{\a} = 0~.
\end{aligned}
\ee
Remembering that $H_{\a}$ is Grassmann odd, its general expression is given by
\bea
H_{\a} = 
\bm{\Q}_{\a} A(\bm{X}) + \bar{\bm \Q}^2 \bm{\Q}_{\a} B(\bm{X}) + \bar{\bm \Q}^{\ad} ({\s}^{m})_{\a \ad} 
C_m (\bm{X}) + \bm{\Q}^2 \bar{\bm \Q}^{\ad} ({\s}^{m})_{\a \ad} D_m (\bm{X})~.
\eea
The second order constraints $\bar \cD^2 H_{\a}= \cQ^2 H_{\a} = 0$ set 
\bea
B = 0~, \qquad D_{m} = 0~.
\eea
Using the property \eqref{hom4}, we can then write
\bea
H_{\a} =  \frac{a_1}{\bm{X}^2} \bm{\Q}_{\a} +  \frac{c_1}{(\bm{X}^2)^{3/2}} \bar{\bm \Q}^{\ad} \bm{X}_{\a \ad}~.
\eea
From here, it is not hard to verify that the requirements $\cD^2 H_{\a} = 0$ and $\bar \cQ^2 H_{\a} = 0$ 
imply  $a_1 = 0$ and $c_1 = 0$, respectively. Thus, we conclude that
\bea
\la L(z_1) L(z_2) S_{\a}(z_3) \ra = 0~.
\eea


\section{Correlators of the spinor current multiplets}  \label{Section5}


For completeness, we will also consider three-point functions of three spinor current multiplets. Such correlators 
carry an $R$-symmetry charge and are expected to vanish. In this section, we will check that it is indeed true.


\subsection{Correlator $\la S_{\a}(z_1) S_{\b}(z_2) S_{\g}(z_3) \ra$}


We begin with the general ansatz
\bea
\la S_{\a}(z_1) S_{\b}(z_2) S_{\g}(z_3) \ra = \frac{(x_{1 \bar{3}})_{\a \ad} (x_{2 \bar{3}})_{\b \dot{\b}}}{x_{\bar{1} 3}{}^2 (x_{\bar{3} 1}{}^2)^2 x_{\bar{2} 3}{}^2 (x_{\bar{3} 2}{}^2)^2 } \, H^{\ad, \, \dot{\b}}\,_{\g}(\fn3)~.
\eea
As a result of imposing \eqref{c1}, the tensor $H_{\ad,\, \g \bd}$ must satisfy the following constraints
\be 
\begin{aligned}
&\bar{\cQ}^{\dot{\b}}H_{\ad,\, \g \dot{\b}} = 0~, \qquad {\cQ}^{2}H_{\ad,\, \g \bd} = 0~,\\
&\bar{\cD}^{\dot{\a}}H_{\ad,\, \g \bd} = 0~, \qquad {\cD^2}H_{\ad,\, \g \bd} = 0~.
\end{aligned}
\label{JJJdiff-con}
\ee
It also obeys the homogeneity property
\bea \label{hom5}
H_{\ad,\, \g \bd}\, (\l \bar{\l} \bm{X}, \l \bm{\Q}, \bar{\l} \bar{\bm \Q} ) = \l^{-3} \bar \l^{-2} H_{\ad,\, \g \bd}\ (\fxq)~.
\eea
Note that we must impose one more condition: under permutation of superspace points $z_1$ and $z_3$, the correlation function must satisfy 
\bea
\la S_{\a}(z_1) S_{\b}(z_2) S_{\g} (z_3) \ra = - \la S_{\g}(z_3) S_{\b}(z_2) S_{\a}(z_1) \ra~.
\eea
However, we will see that the constraints \eqref{JJJdiff-con} and  \eqref{hom5} are sufficient to show that the correlator vanishes. 

Let us trade a pair of spinor indices with the vector one, i.e. 
\bea
H_{\ad,\, \g \bd} = (\s^m)_{\g \dot{\b}} H_{m,\, \ad}~,
\eea
where $H_{m,\, \ad} (\fxq)$ being Grassmann odd has the following expansion
\be
\begin{aligned}
H_{m,\, \ad} (\fxq) &=  \bm{\Q}^{\d} (\s^n)_{\d \ad} A_{mn} (\bm{X}) + \bar{\bm \Q}^2 \bm{\Q}^{\d} (\s^n)_{\d \ad} B_{mn} (\bm{X})\\ 
&+\bar{\bm \Q}^{\dot{\d}} C_{m, \,\dot{\d}, \ad} (\bm{X}) + \bm{\Q}^2 \bar{\bm \Q}^{\dot{\d}} D_{m, \,\dot{\d}, \ad} (\bm{X})~.
\end{aligned}
\ee
One can readily verify that the conditions ${\cQ}^{2}H_{\ad,\, \g \bd} = 0$ and $\bar{\cD}^{\dot{\a}}H_{\ad,\, \g \bd} = 0$ give 
\bsubeq
\bea
&&B_{mn} = 0~, \qquad  D_{m, \,\dot{\d}, \ad} = 0~,\\
&&\ve^{\ad \dot{\d}}C_{m, \,\dot{\d}, \ad} = 0~.
\eea
\esubeq
Thus, it suffices for us to determine the general ansatz for $A_{m, n}$ and $C_{m,\, (\dot{\d} \ad)}$. In order to be compatible with \eqref{hom5}, we should write
\bsubeq
\bea
&& A_{mn} = a_1 \frac{\eta_{mn}}{(\bm{X}^2)^{3/2}} + a_2 \frac{\bm{X}_m \bm{X}_n}{(\bm{X}^2)^{5/2}}~,\\
&& C_{m,\, (\dot{\d} \ad)} = (\tilde{\s}^{nk})_{\dot{\d} \ad} 
\bigg [ \frac{c_1}{(\bm{X}^2)^2} (\eta_{mn} \bm{X}_k - \eta_{mk} \bm{X}_n) + \frac{c_2}{(\bm{X}^2)^2} \e_{mnkp} \bm{X}^p \bigg]~.
\eea
\esubeq
However, when the remaining constraints $\bar{\cQ}^{\dot{\b}}H_{\ad,\, \g \bd} = 0$ and ${\cD^2}H_{\ad,\, \g \bd}= 0$ are taken into account, we find that
$a_1= a_2=c_1= c_2=0$ and, thus, 
\bea
\la S_{\a}(z_1) S_{\b}(z_2) S_{\g}(z_3) \ra = 0~.
\eea


\subsection{Correlator $\la S_{\a}(z_1) \bar S_{\bd}(z_2) S_{\g}(z_3) \ra$}


We again begin with the general ansatz
\bea
\la S_{\a}(z_1) \bar S_{\bd}(z_2) S_{\g}(z_3) \ra = \frac{(x_{1 \bar{3}})_{\a \ad} (x_{3 \bar{2}})_{\b \dot{\b}}}{x_{\bar{1} 3}{}^2 (x_{\bar{3} 1}{}^2)^2 (x_{\bar{2} 3}{}^2)^2 x_{\bar{3} 2}{}^2 } \, H^{\ad, \, \b}\,_{\g}(\fn3)~.
\eea
As a result of imposing \eqref{c1}, the tensor $H_{\b, \, \g \ad}$ must satisfy the following constraints
\be
\begin{aligned}
&{\cQ}^{{\b}}H_{\b, \, \g \ad} = 0~, \qquad \bar{\cQ}^{2}H_{\b, \, \g \ad} = 0~,\\
&\bar{\cD}^{\dot{\a}}H_{\b, \, \g \ad} = 0~, \qquad {\cD^2}H_{\b, \, \g \ad} = 0~.
\end{aligned}
\ee
It also obeys the homogeneity property
\bea \label{hom6}
H_{\b,\, \g \ad}\, (\l \bar{\l} \bm{X}, \l \bm{\Q}, \bar{\l} \bar{\bm \Q} ) = \l^{-8/3} \bar \l^{-7/3} H_{\b,\, \g \ad}\ (\fxq)~.
\eea

Let us trade a pair of spinor indices with the vector one, i.e. 
\bea
H_{\b, \, \g \ad} = (\s^m)_{\g \dot{\a}} H_{m,\, \b}~,
\eea
where $H_{m,\, \b} (\fxq) $ being Grassmann odd has the following expansion
\be
\begin{aligned}
H_{m,\, \b} (\fxq) &=  \bm{\Q}^{\d} A_{m, \, \d,\, \b} (\bm{X}) + \bm{\bar \Q}^2 \bm{\Q}^{\d} B_{m,\, \d, \,\b} (\bm{X})\\ 
&+\bm{\bar \Q}^{\dot{\d}} (\s^n)_{\b \dot{\d}}\, C_{mn} (\bm{X}) + \bm{\Q}^2 \bm{\bar \Q}^{\dot{\d}} (\s^n)_{\b \dot{\d}} \,D_{mn} (\bm{X})~.
\end{aligned}
\ee
One can readily verify that the conditions $\bar{\cD}^{\dot{\a}}H_{\b, \, \g \ad} = 0$ and ${\cQ}^{{\b}}H_{\b, \, \g \ad} = 0$ give 
\bsubeq
\bea
&&B_{m,\, \d, \,\b} = 0~, \qquad  D_{mn} = 0~, \qquad 
\ve^{\b \d}A_{m, \, \d,\, \b}  = 0~, \\
&&\eta^{mn} C_{mn} = 0~, \qquad C_{[mn]} = 0~. \label{corr6-alg3}
\eea
\esubeq
Thus, it suffices for us to determine the general ansatz for $A_{m,\, ({\d} \b)}$ and $C_{(mn)}$ only. In order to be compatible with \eqref{hom6} and \eqref{corr6-alg3}, we should write
\bsubeq
\bea
 A_{m,\, ({\d} \b)} &=& (\tilde{\s}^{nk})_{{\d} \b} 
\bigg [ \frac{a_1}{(\bm{X}^2)^2} (\eta_{mn} \bm{X}_k - \eta_{mk} \bm{X}_n) + \frac{a_2}{(\bm{X}^2)^2} \e_{mnkp} \bm{X}^p \bigg]~, \\
 C_{(mn)} &=& c_1 \bigg[ \frac{\eta_{mn}}{(\bm{X}^2)^{3/2}} -4 \frac{\bm{X}_m \bm{X}_n}{(\bm{X}^2)^{5/2}} \bigg]~.
\eea
\esubeq
However, it turns out that the second order constraints ${\cD^2}H_{\b, \, \g \ad} = 0$ and $\bar{\cQ}^{2}H_{\b, \, \g \ad} = 0$ are enough to show 
that $a_1= a_2= c_1=0$. Thus, 
\bea
\la S_{\a}(z_1) \bar{S}_{\bd}(z_2) S_{\g}(z_3) \ra = 0~.
\eea


\section{${\cN}=2 \to {\cN}=1$ superspace reduction} \label{Section6}

In this section we will discuss whether the three-point functions studied in the previous sections are consistent with $\cN=2$ superconformal 
symmetry. For this we will study if they are contained in the three-point function of the $\cN=2$ supercurrent. 


\subsection{The three-point function of the $\cN=2$ supercurrent}


Let us recall that in ${\cN}=2$ superconformal field theory, the supercurrent \cite{Sohnius:1978pk, HST} is described by a 
primary real scalar superfield $\cJ(z)$ of dimension 2, subject to the conservation condition
\bea
D^{ij} \cJ = 0 \quad \Longleftrightarrow \quad \bar D^{ij} \cJ = 0~.
\label{6.1}
\eea
Here $D_{A} = (\pa_a, D_{\a}^i, \bar{D}_{i}^{\ad})$ are the ${\cN}=2$ supersymmetric covariant derivatives, 
$i = \1, \2$, and $D^{ij} = D^{\a(i} D_{\a}^{j)}, \, \bar D^{ij} = \bar{D}_{\ad}^{(i} \bar{D}^{j) \ad}$.
The superconformal transformation law of $\cJ$, which is uniquely determined from the reality condition $\bar{\cJ} = \cJ$ 
and the conservation equation, reads
\bea
\d \cJ(z) = -\xi \cJ(z) -2 (\s(z) + \bar{\s} (z)) \cJ(z)~.
\label{n=2sctrl}
\eea
As was discussed in the introduction, 
the ${\cN}=2$ supercurrent $\cJ$ is composed of three independent ${\cN}=1$ multiplets \cite{KT}. For convenience, we will write them again
\bea
J \equiv \cJ| = \bar{J}~, \quad J_{\a} \equiv D^{\2}_{\a} \cJ|~, \quad 
J_{\a \ad} \equiv \hf [D^{\2}_{\a}, \bar{D}_{\ad \2}] \cJ| - \frac{1}{6} [D^{\1}_{\a}, \bar{D}_{\ad \1}] \cJ| = \bar{J}_{\a \ad}~.
\label{6.2}
\eea
It follows from eqs.~\eqref{6.1} and~\eqref{6.2} that the $\cN=1$ components of $\cJ$ satisfy the following conservation equations
\bsubeq  \label{N1comp-supercurrent}
\bea
 \bar{D}^2 J &=& 0~, \qquad D^2 J =0\\
D^{\a}J_{\a} &=& 0~, \qquad \bar{D}^2 J_{\a} = 0~,\\
\bar{D}^{\ad}J_{\a \ad} &=& 0~, \qquad  D^{\a}J_{\a \ad} = 0~.
\eea
\esubeq
From eqs.~\eqref{n=2sctrl} and~\eqref{6.2} one can also deduce the transformation properties
\bsubeq
\label{6.3}
\bea
\d J  &=& -\x\,J    
 - 2\left(  \s  +  \bar{\s}  \right) J  \label{n=1trf-J}\\
\d J_\a  &=& - \x\,J_\a  +\hat{\o}_\a{}^\b 
 J_\b  - \left(3 
\s +  2 \bar{\s}  \right) J_\a  \\ 
\d J_{\a \ad}  &=& - \x\,J_{\a \ad} 
+ ( \hat{\o}_\a{}^\b  \d_\ad{}^\bd 
+ \bar{\hat{\o}}_\ad{}^\bd  \d_\a{}^\b) J_{\b \bd} 
- 3\left(  
\s +  \bar{\s}  \right) J_{\a \ad} ~.
\eea
\esubeq
The above equations~\eqref{N1comp-supercurrent} and~\eqref{6.3} 
imply that ${J}_{\a \ad}$ is the ${\cN}=1$ supercurrent, $J_{\a}$ is a spinor current and $J$ is a flavour-like 
current corresponding to one of the $U(1)$ generators of the $R$-symmetry. 

If an $\cN=1$ superconformal theory also has $\cN=2$ superconformal symmetry, all three-point functions 
involving any combination of $J, J_{\a}, J_{\a \ad}$ must then be contained in the three-point function of the $\cN=2$ supercurrent. 
More precisely, they can be found from $\la \cJ(z_1) \cJ(z_2) \cJ(z_3) \ra$ by making use of the ${\cN}=2 \to {\cN}=1$ superspace reduction formalism of \cite{KT}.
Hence, our aim will be to consider the relevant superspace reductions of $\la \cJ(z_1) \cJ(z_2) \cJ(z_3) \ra$ to see if we can reproduce the results for $\la \bar S_{\ad}(z_1) S_{\b}(z_2) J_{\g \gd}(z_3) \ra$
and $\la \bar S_{\ad}(z_1) S_{\b}(z_2) L(z_3) \ra$ obtained in the previous sections.

The three-point function of $\cN=2$ supercurrent was found in~\cite{KT}. It is given by 
\bea \label{N2-3pt-J}
\la \cJ(z_1) \cJ(z_2) \cJ(z_3) \ra &=& \frac{1}{x_{\bar{1} 3}{}^2 x_{\bar{3} 1}{}^2 x_{\bar{2} 3}{}^2 x_{\bar{3} 2}{}^2 } H(\bm{Z}_3)~,\non\\
H(\l \bar{\l} \bm{X}, \l \bm{\Q}, \bar{\l} \bar{\bm{\Q}}) &=& (\l \bar{\l})^{-2} H(\fxq)~, 
\eea
where the real function $H(\fxq)$ is
\bea \label{N2-H}
H(\fn3) &=& A \Big( \frac{1}{\bm{X}_3^2} + \frac{1}{\bm{\bar X}_3^2}\Big) + B \frac{ \bm{\Q}_3^{\a \b} \bm{X}_{3 \a \ad} \bm{X}_{3 \b \bd} \bm{\bar \Q}_3^{\ad \bd} }{(\bm{X}_3^2)^3} \non\\
&=& \Big(A+ \frac{B}{8}\Big) \bigg( \frac{1}{\bm{X}_3^2} + \frac{1}{\bm{\bar X}_3^2}\bigg) - \frac{B}{8} (\bm{\bar X}_3 \cdot \bm{X}_3 ) \Big(\frac{1}{(\bm{X}_3^2)^2} + \frac{1}{(\bm{\bar X}_3^2)^2} \Big)~. 
\eea
In the above, $A$ and $B$ are arbitrary real parameters and
\bea
\bm{\Q}_{3}^{\a \b} = \bm{\Q}_{3}^{(\a \b)} = \bm{\Q}_{3}^{\a i} \bm{\Q}_{3 i}^{\b}~, \qquad \bm{\bar \Q}_{3}^{\ad \bd} = \bm{\bar \Q}_{3}^{(\ad \bd)} = \bm{\bar \Q}_{3 i}^{\ad} \bm{\Q}_{3}^{\bd i}~.
\eea
The second line of \eqref{N2-H} is obtained with the aid of the following identities~\cite{KT}:
\bea
&&
\frac{1}{\bm{\bar X}_3^2} = \frac{1}{\bm{X}_3^2} - 4 \ri \, \frac{\bm{\Q}_3^i \bm{X}_3 \bm{\bar \Q}_{3i}}{(\bm{X}_3^2)^2} 
-8 \frac{ \bm{\Q}_3^{\a \b} \bm{X}_{3 \a \ad} \bm{X}_{3 \b \bd} \bm{\bar \Q}_3^{\ad \bd} }{(\bm{X}_3^2)^3}~,
\non \\
&&
\bm{\bar X}^{a}_3 = \bm{X}_3^{a} + 2 \ri \bm{\Q}^i_{3} \s^a \bm{\bar {\Q}}_{3i}~.
\label{6.4}
\eea
%
In the next subsections we will extract 
$\la \bar J_{\ad}(z_1) J_{\b}(z_2) J_{\g \gd}(z_3) \ra$
and $\la \bar J_{\ad}(z_1) J_{\b}(z_2) J(z_3) \ra$ by superspace reduction and compare the results with eq.~\eqref{corr1fin} 
and~\eqref{tensor-JJL} respectively. 


\subsection{Correlator $\la \bar J_{\ad} (z_1) J_{\b}(z_2) J_{\g \gd}(z_3) \ra$}


Extracting three-point functions involving the ${\cN}=1$ supercurrent $J_{\a \ad}$ is technically quite difficult. 
Following similar considerations in~\cite{KT}, we define the operator
\bea
\D_{\a \ad}= \hf [D^{\2}_{\a}, \bar{D}_{\ad \2}]  - \frac{1}{6} [D^{\1}_{\a}, \bar{D}_{\ad \1}]~,
\eea
and hence $J_{\a \ad}$ is obtained from $\cJ$ by applying the operator $\D_{\a\ad}$ and then switching off the Grassmann variables $\q_{\2}, \bar{\q}^{\2}$. 
As a first step, let us consider $\la \D_{\a \ad} \cJ(z_1) \cJ(z_2) \cJ(z_3) \ra$. 
In~\cite{KT} it was found that
\be
\begin{aligned}
&\la \D_{\a \ad} \cJ(z_1) \cJ(z_2) \cJ(z_3) \ra = \frac{(\tilde{x}^{-1}_{\bar{3}1})_{\a \dot{\s}} (\tilde{x}^{-1}_{\bar{1} 3})_{\s \ad}}{x_{\bar{1} 3}{}^2 x_{\bar{3} 1}{}^2 x_{\bar{2} 3}{}^2 x_{\bar{3} 2}{}^2 }\\
&\qquad \quad \times \Big( \frac{64}{3} \q^{\s}_{13 \2} \,\bar{\q}^{\dot{\s} \2}_{13}\, H (\bm{Z}_3) + \frac{16}{3} \q^{\s}_{13 \2} \,u_k^{\2}(z_{31}) \bar{\cD}^{\dot{\s} k} H (\bm{Z}_3)- \frac{16}{3} {\bar{\q}}^{\dot{\s} \2}_{13} \,u_{\2}^{k}(z_{13}) {\cD}^{{\s} k} H (\bm{Z}_3)\\
&\qquad \quad + \hf \Big\{ u_{\2}^{k}(z_{13}) \,u_{l}^{\2}(z_{31}) - \frac{1}{3} u_{\1}^{k}(z_{13}) \,u_{l}^{\1}(z_{31}) \Big\} [\cD^{\s}_{k}, \bar{\cD}^{\dot{\s} l}] H(\bm{Z}_3) \Big)~.
\end{aligned}
\ee
Next, we may express $H(\bm{Z}_3), \cD^{\s}_{k} H(\bm{Z}_3)$ and $[\cD^{\s}_{k}, \bar{\cD}^{\dot{\s} l}] H(\bm{Z}_3)$ as functions of $\bm{Z}_1$, by virtue of \eqref{Z13}. Upon relabelling $z_1 \leftrightarrow z_3$ (and noting the relation \eqref{pt13}), we then obtain
\bea
\la  \cJ(z_1) \cJ(z_2) \,\D_{\g \gd}\cJ(z_3) \ra~,
\eea
which is a function of $\bm Z_3$. 
Thus, the three-point function \eqref{corr1} can be derived by evaluating the bar-projection
\bea
\la \bar{J}_{\ad} (z_1) J_{\b}(z_2) J_{\g \gd}(z_3)\ra = \bar{D}_{(1) \ad \2} \,D^{\2}_{(2) \b} \la \cJ(z_1) \cJ(z_2) \D_{\g \gd}\cJ(z_3) \ra \big|~,
\eea
and making use of the identities \eqref{N2covderiv}. 

After quite a long calculation, the above procedure yields
\bea
\la \bar{J}_{\ad}(z_1) J_{\b}(z_2) J_{\g \gd} (z_3)\ra = \frac{(x_{3 \bar{1}})_{\a \ad} (x_{2 \bar{3}})_{\b \bd}}{ (x_{\bar{1} 3}{}^2)^2  (x_{\bar{3} 2}{}^2)^2 x_{\bar{3} 1}{}^2 x_{\bar{2} 3}{}^2 } {\cH}^{\bd \a}\,_{\g \gd} (\fn3)~,
\label{6.5}
\eea
where
\bea
&&{\cH}^{\bd \a}\,_{\g \gd} (\bm{X}, \bm{\Q}, \bm{\bar \Q}) = \frac{16}{3}\Big(A-\frac{3B}{8}\Big) \frac{1}{\bm{X}^2} \d^{\a}_{\g} \d^{\bd}_{\gd}- \frac{16}{3}\Big(A+\frac{B}{8}\Big) \frac{1}{(\bm{X}^2)^2} \bm{X}^{\bd \a} \bm{X}_{\g \gd} \non\\
&&\qquad \quad - \frac{16}{3}\Big(A - \frac{3B}{8}\Big) \frac{\ri}{(\bm{X}^2)^2} \bm{X}^{\bd \a} \bm{P}_{\g \gd} - \frac{8}{3} B \frac{\ri}{(\bm{X}^2)^2} \bm{P}^{\bd \a} \bm{X}_{\g \gd} \non\\
&&\qquad \quad - \frac{32}{3} \Big( A + \frac{B}{8} \Big) \frac{\ri}{(\bm{X}^2)^2} (\bm{P} \cdot \bm{X}) \d^{\a}_{\g} \d^{\bd}_{\gd} + \frac{16}{3}B \frac{\ri}{(\bm{X}^2)^3} (\bm{P} \cdot \bm{X}) \bm{X}^{\bd \a} \bm{X}_{\g \gd}~.
\label{6.6}
\eea
Thus, any $\cN=1$ superconformal field theory possessing a conserved spinor current multiplet $S_{\a}$ also has $\cN=2$ superconformal
symmetry, if and only if the correlator $\la \bar{S}_{\ad} (z_1) S_{\b}(z_2) J_{\g \gd}(z_3)\ra$  is of the form~\eqref{6.5}, \eqref{6.6}. 
Comparing the above with \eqref{corr1fin}, we see that they agree provided
\bea
d_2 = \frac{32}{3} \Big(A + \frac{B}{8}\Big)~, \qquad d_3 = -\frac{16}{3}B~, \qquad d_4 = -\frac{32}{3} \Big(A- \frac{3B}{8}\Big)~.
\label{6.7}
\eea
This implies 
\bea
d_2 + d_3 + d_4 = 0~.
\label{6.8}
\eea
Therefore, we obtain the following surprising result: Existence of a conserved spinor current in $\cN=1$ superconformal 
field theory, in general, does not imply $\cN=2$ superconformal symmetry. Only if the coefficients $d_2, d_3, d_4$ in~\eqref{corr1fin}
satisfy the additional constraint~\eqref{6.8} the three-point function~\eqref{corr1fin} is contained in the three-point 
function of the $\cN=2$ supercurrent and can be obtained by the superspace reduction. Otherwise, eq.~\eqref{corr1fin}
contains an extra independent tensor structure incompatible with $\cN=2$ superconformal symmetry. 
More precisely, the tensor $H_{mn}$ in \eqref{corr1fin} may be expressed in the following way
\bea
H_{mn} (\bm{X}, \bm{\Q}, \bar{\bm{\Q}}) = \cH_{mn} (\bm{X}, \bm{\Q}, \bar{\bm{\Q}}) + \hf (d_2+ d_3+ d_4) \frac{1}{(\bm X^2)^2}\Big\{\bm{X}_m \bm{X}_n + \ri \bm{X}_m \bm{P}_n \Big\}~,~~~
\eea
where $\cH_{mn}$ is eq. \eqref{6.6} written in vector notation, that is
\bea
&&\cH_{mn} (\bm{X}, \bm{\Q}, \bar{\bm{\Q}}) = 
-\frac{1}{4}(d_2+d_3) \frac{\eta_{mn}}{\bm{X}^2} - \hf d_2 \frac{\bm{X}_m \bm{X}_n}{(\bm{X}^2)^2} \non\\
&&\quad \quad - \hf (d_2+d_3) \frac{\ri}{(\bm{X}^2)^2} \bm{X}_m \bm{P}_n +  \hf d_3 \frac{\ri}{(\bm{X}^2)^2} \bm{X}_n \bm{P}_m \non\\
&&\quad \quad + \hf d_2 \frac{\ri}{(\bm{X}^2)^2} \eta_{mn}(\bm{P} \cdot \bm{X} ) - d_3 \frac{\ri}{(\bm{X}^2)^3} (\bm{P} \cdot \bm{X}) \bm{X}_m \bm{X}_n~.
\eea
In section 7, we will present two consistency checks of our result \eqref{corr1fin}. 


\subsection{Correlator $\la \bar{J}_{\ad}(z_1) J_{\b}(z_2) J(z_3) \ra$}


In this subsection, we are going to show that the three-point function $\la \bar{S}_{\ad}(z_1) S_{\b}(z_2) J(z_3) \ra$, eq. \eqref{JJL-fin}, is consistent with $\cN=2$ superconformal symmetry and must come from the 
superspace reduction of $\la \cJ(z_1) \cJ(z_2) \cJ(z_3) \ra$.

It follows from eqs.~\eqref{6.2} that in $\cN=2$ superconformal theory, the three-point function $\la \bar{J}_{\ad}(z_1) J_{\b}(z_2) J(z_3) \ra$
can be extracted using the following superspace reduction:
\bea
\la \bar{J}_{\ad}(z_1) J_{\b}(z_2) J(z_3) \ra = \bar D_{(1) \ad \2} D^{\2}{}_{(2)\b} \la \cJ(z_1) \cJ(z_2) \cJ(z_3)\ra \big|~.
\label{6.9}
\eea
Keeping in mind that
\bea
D_{(2) \b}^{i} \bm{\bar X}_{3 \g \gd} = 0~, \qquad \bar{D}_{(1) \ad\, i} \bm{\bar X}_{3 \g \gd} = 0~,
\eea
along with the identities~\eqref{N2covderiv}, one obtains
\bea
\la \bar{J}_{\ad}(z_1) J_{\b}(z_2) J(z_3) \ra = \frac{(x_{3 \bar{1}})_{\a \ad} (x_{2 \bar{3}})_{\b \bd}}{(x_{\bar{1} 3}{}^2)^{2} x_{\bar{3}1}{}^2 x_{\bar{2} 3}{}^2(x_{\bar{3} 2}{}^2)^{2}} {\cH}^{\bd \a}(\fn3)~,
\label{6.10}
\eea
with 
\bea
\cH^{\bd \a}(\fxq) = -4\ri\, A \frac{\bm X^{\bd \a}}{(\bm X^2)^2}
+ \frac{B}{2 (\bm X^2)^3} \Big( \bm X^2 \bm P^{\bd \a} - 4 (\bm P \cdot \bm X) \bm X^{\bd \a} \Big)~.
\label{N2red-JJL}
\eea
Comparing eqs.~\eqref{6.10}, \eqref{N2red-JJL} with the ${\cN}=1$ results~\eqref{JJL-fin}, \eqref{tensor-JJL}
we see that they are in perfect agreement, provided the coefficients are related via
\be
c_1 = -4 A~, \qquad d_1 = B~.
\ee
That is, the $\cN=1$ correlator $\la \bar{S}_{\ad}(z_1) S_{\b}(z_2) J(z_3) \ra$ is always consistent with $\cN=2$ superconformal symmetry. 


\subsection{Correlator $\la \bar{J}_{\ad}(z_1) J_{\b}(z_2) L(z_3) \ra$}
The ${\cN}=2$ flavour current superfield \cite{KT} is subject to the reality condition $\overline{\cL_{ij} } =
\cL^{ij} $, and the conservation (analyticity) equation
\be
D^{(i}_\a \cL^{jk)} = {\bar D}^{(i}_\ad \cL^{jk)} =0~.
\label{fcce}
\ee
These constraints fix its transformation law to be
\be
\d \cL_{ij}(z) = -\x \, \cL_{ij}(z) 
+ 2 {\rm i}\, \hat{\L}_{(i}{}^k (z)\, \cL_{j)k}(z)
- 2\left(  \s(z) +  \bar{\s} (z) \right)\cL_{ij}(z)~.
\label{n=2fcstl}
\ee  
Its leading $\cN=1$ component 
\be
L ~ \equiv ~{\rm i}\, \cL^{ \underline{12} }| ~=~\bar{L}
\label{n=1fcs}
\ee
obeys the conservation equation
\be
D^2\, L ~=~\bar{D}^2\, L ~=~0~.
\ee
Hence, its superconformal transformation 
rule is similar to $J$ (see eq. \eqref{n=1trf-J})
\be
\d L  = -\x\,L    
 - 2\left(  \s  +  \bar{\s}  \right) L~. 
\ee 

It was demonstrated in \cite{KT} that the three-point function involving two $\cN=2$ supercurrent
insertions and a flavour $\cN=2$ current superfield vanishes,
\be
\langle
\cJ (z_1) \; \cJ (z_2) \; \cL_{ij} (z_3)
\rangle ~=~ 0\;.
\ee
As a result, 
\bea
\la \bar{J}_{\ad}(z_1) J_{\b}(z_2) L(z_3) \ra = - \ri \, \bar D_{(1) \ad \2} D^{\2}{}_{(2)\b} \, \la \cJ(z_1) \cJ(z_2) \cL_{\2 \1} (z_3) \ra \big| = 0~.
\eea 


\section{Reduction to component currents}
\label{Section7}


In ${\cN}=1$ superconformal field theory, the spinor current $S_{\a}$ and the supercurrent $J_{\a \ad}$ multiplets contain, among others, 
the following conserved currents:
\bea
&&\mathbb{V}_{\a \ad} := \bar{D}_{\ad} S_{\a} |~,
\label{77.1}\\
&&R_{\a \ad} := J_{\a \ad} |~, \qquad T_{\a \b \ad \bd} := [D_{(\b}, \bar{D}_{(\bd}] J_{\a )\ad)}|~.
\label{77.2}
\eea
Here the bar-projection corresponds to setting the Grassmann coordinates to zero.
The vector current $\mathbb{V}_{\a \ad}$ is a complex field,
\be
\mathbb{V}_{\a \ad} = V_{\a \ad} + \ri \tilde{V}_{\a \ad}~.
\label{77.3}
\ee
The components $R_{\a \ad}$ and $T_{\a\b \ad \bd}$ are the $\rm U(1)$ $R$-symmetry current and the energy-momentum tensor, respectively.
This means that the correlator 
\be 
\la \bar S_{\ad}(z_1) S_{\b}(z_2) J_{\g \gd}(z_3) \ra
\label{77.4}
\ee
in particular contains the following component correlators
\be 
\la V_{\a \ad}(x_1) V_{\b \bd}(x_2) R_{\g \gd}(x_3) \ra~, \qquad 
\la V_{\b \bd}(x_1) T_{\g \d \gd \dot{\d}} (x_2) V_{\a \ad}(x_3) \ra~. 
\label{77.5}
\ee
The three-point functions of this kind were studied in~\cite{OP, EO}.\footnote{Three-point functions involving $R$-symmetry currents 
and various primary operators were recently studied in~\cite{Manenti:2018xns}.} It was shown that the three-point function of two $\rm U(1)$ 
vector currents and an axial current
has only one independent structure, and the three-point function of two vector currents and the energy-momentum tensor 
is fixed up to two independent structures. On the other hand, the correlator~\eqref{77.4} was shown in subsection \ref{Subsec3.1} to depend 
on three linearly independent structures. Hence, it is not trivial to see if our results for the correlator~\eqref{77.4} are consistent with 
the general structure of the correlators~\eqref{77.5} derived in~\cite{OP, EO}.
In this section we will perform a component reduction of the superfield correlator~\eqref{77.4}
to extract the component correlators~\eqref{77.5}, thus providing consistency checks of our results.


\subsection{Correlator $\la V_{\a \ad}(x_1) V_{\b \bd}(x_2) R_{\g \gd}(x_3) \ra$}


Our first task is to extract the correlator $\la V_{\a \ad}(z_1) V_{\b \bd}(z_2) R_{\g \gd}(z_3) \ra$ from \eqref{corr1fin}, 
where ${V}_{\a \ad}$ denotes the real part of $\mathbb{V}_{\a \ad}$,
\bea
V_{\a \ad} = {\rm Re}(\mathbb{V}_{\a \ad}) = \hf (\bar{D}_{\ad} S_{\a}- D_{\a} \bar{S}_{\ad})|~.
\eea
The correlation function can then be computed using
\bea
\la V_{\a \ad} (x_1) V_{\b \bd} (x_2) R_{\g \gd} (x_3)\ra =  \frac{1}{4} \Big( D_{(1)\a} \bar{D}_{(2)\bd}\, \la \bar{S}_{\ad} (z_1) S_{\b} (z_2) J_{\g \gd} (z_3)\ra\, \big| + \rm c.c. \Big)~.
\eea
In the above, we have used the fact that the correlators $\la \bar{S} \bar{S} J_{\g \gd} \ra = \la S S J_{\g \gd} \ra$ are vanishing, see subsection \ref{ss32}. 
With the help of identities \eqref{Cderivs}, it is easy to see that
\bea
&&D_{(1)\a} \bar{D}_{(2) \bd} \la \bar{S}_{\ad} (z_1) S_{\b} (z_2) J_{\g \gd} (z_3)\ra | = \frac{(x_{1 \bar{3}})_{\a \dot{\m}} (x_{3 \bar{1}})_{\d \ad} (x_{3 \bar{2}})_{\m \bd} (x_{2 \bar{3}})_{\b \dot{\d}} }{(x_{\bar{1} 3}{}^2)^3 (x_{\bar{3} 2}{}^2)^3 x_{\bar{3} 1}{}^2 x_{\bar{2} 3}{}^2} \bar \cD^{\dot{\m}} \cQ^{\m} H^{\dot{\d} \d}{}_{\g \gd}\, \big|~\\
&&= - \frac{(x_{13})_{\a \dot{\m}} (x_{13})_{\d \ad} (x_{23})_{\m \bd} (x_{23})_{\b \dot{\d}} }{(x_{13}{}^2 x_{23}{}^2)^4} \frac{\pa}{\pa \bar{\bm \Q}_{\dot{\m}}} \frac{\pa}{\pa \bm{\Q}_{\m}}  H^{\dot{\d} \d}{}_{\g \gd}\, \big|~,
\eea
and hence, after taking complex conjugation into account,
\bea
&&\la V_{\a \ad} (x_1) V_{\b \bd} (x_2) R_{\g \gd} (x_3)\ra \non\\
&&= -\frac{1}{4}\frac{(x_{13})_{\a \dot{\m}} (x_{13})_{\d \ad} (x_{23})_{\m \bd} (x_{23})_{\b \dot{\d}} }{(x_{13}{}^2 x_{23}{}^2)^4} \bigg\{ \frac{\pa}{\pa \bar{\bm \Q}_{\dot{\m}}} \frac{\pa}{\pa \bm{\Q}_{\m}}  H^{\dot{\d} \d}{}_{\g \gd}\, \big|  - \frac{\pa}{\pa \bar{\bm \Q}_{\dot{\d}}} \frac{\pa}{\pa \bm{\Q}_{\d}}  H^{\dot{\m} \m}{}_{\g \gd}\, \big| \bigg\}~. 
\label{eq710}
\eea
Here we have defined $x_{1 \bar{3}} | = x_{13} = x_1 - x_3 = -x_{31}$. We find that 
\bea
&&\frac{\pa}{\pa \bar{\bm \Q}_{\dot{\m}}} \frac{\pa}{\pa \bm{\Q}_{\m}}  H^{\dot{\d} \d}{}_{\g \gd}\, \big| = -2 \ri \bigg\{ d_4 \,\d^{\m}{}_{\g} \d^{\dot{\m}}{}_{\gd} \frac{X^{\dot{\d} \d}}{X^4} + d_2 \,\d^{\d}{}_{\g} \d^{\dot{\d}}{}_{\gd} \frac{X^{\dot{\m} \m}}{X^4} \non\\
 &&\qquad \qquad \qquad \quad + d_3 \Big( \ve^{\m \d} \ve^{\dot{\m} \dot{\d}} \frac{X_{\g \gd}}{X^4} + \frac{X^{\dot{\d} \d} X^{\dot{\m} \m} X_{\g \gd}}{X^6}\Big) \bigg\}~.
\eea
It follows from \eqref{eq710} that there is just one linearly independent form for the correlator,
\bsubeq 
\bea
\la V_{\a \ad} (x_1) V_{\b \bd} (x_2) R_{\g \gd} (x_3)\ra = \frac{\ri}{2} (d_4 - d_2) \frac{(x_{13})_{\a \dot{\m}} (x_{13})_{\d \ad} (x_{23})_{\m \bd} (x_{23})_{\b \dot{\d}} }{(x_{13}{}^2 x_{23}{}^2)^4} t^{\dot{\m} \m, \,\dot{\d} \d}{}_{\g \gd} (X_3)~,~~~~
\eea
where the function $t^{\dot{\m} \m, \,\dot{\d} \d}{}_{\g \gd} (X)$ takes the form
\bea
t^{\dot{\m} \m,\, \dot{\d} \d}{}_{\g \gd} (X) = \d^{\m}{}_{\g} \d^{\dot{\m}}{}_{\gd} \frac{X^{\dot{\d} \d}}{X^4} - 
\d^{\d}{}_{\g} \d^{\dot{\d}}{}_{\gd} \frac{X^{\dot{\m} \m}}{X^4}~,
\eea
\esubeq
with $X_{\a \ad} = \bm{X}_{\a \ad}| = \bar{\bm X}_{\a \ad}|$~. 
Hence, the above correlator indeed contains a single structure in agreement with ref.~\cite{EO}. 


\subsection{Correlator $\la V_{\b \bd}(x_1) T_{\g \d \gd \dot{\d}} (x_2) V_{\a \ad}(x_3) \ra$}


To derive the three-point function involving two vector currents and the energy-momentum tensor from \eqref{corr1fin} 
it is convenient to start with the following equivalent expression 
\bea
\la S_{\b} (z_1) J_{\g \gd}(z_2) \bar{S}_{\ad} (z_3) \ra = \frac{(x_{1 \bar{3}})_{\b \dot{\m}} (x_{2 \bar{3}})_{\g \dot{\l}} (x_{3 \bar{2}})_{\l \gd} }{x_{\bar{1} 3}{}^2 (x_{\bar{3} 1}{}^2)^2 (x_{\bar{2} 3}{}^2)^2 (x_{\bar{3} 2}{}^2)^2 } \tilde{H}_{\ad}^{\dot{\m}, \, \dot{\l} \l} (\fn3)~,
\label{SJS-1}
\eea
where the explicit form of $\tilde{H}_{\ad}^{\dot{\m}, \, \dot{\l} \l}$ is given by \eqref{326}. In order to obtain the correlation function $\la V_{\b \bd}(x_1) \,T_{\g \d \gd \dot{\d}} (x_2) \, V_{\a \ad}(x_3) \ra$, we must act with the following derivatives on \eqref{SJS-1}:
\bea
&&\la V_{\b \bd}(x_1) T_{\g \d \gd \dot{\d}} (x_2) V_{\a \ad}(x_3) \ra  \non\\
&&=  -\frac{1}{4} D_{(3) \a} \bar{D}_{(1) \bd} [D_{(2)\, (\d}, \bar{D}_{(2) \,(\dot{\d}}] \la S_{\b} (z_1) J_{\g) \gd)}(z_2) \bar{S}_{\ad} (z_3) \ra \big| + \rm c.c.~.
\eea
Let us first compute
\bea
&&[D_{(2)\, (\d}, \bar{D}_{(2) \,(\dot{\d}}] \, \la S_{\b} (z_1) J_{\g) \gd)}(z_2) \bar{S}_{\ad} (z_3) \ra = \frac{(x_{1 \bar{3}})_{\b \dot{\m}} (x_{2 \bar{3}})_{(\g \dot{\l}} (x_{2 \bar{3}})_{\d) \dot{\rho}} (x_{3 \bar{2}})_{\l (\gd} (x_{3 \bar{2}})_{\rho \dot{\d})} }{x_{\bar{1} 3}{}^2 (x_{\bar{3} 1}{}^2)^2 (x_{\bar{2} 3}{}^2)^3 (x_{\bar{3} 2}{}^2)^3 }\non\\
&&\qquad \qquad \times \Big( 160 \,\q^{\rho}_{23} \,\bar{\q}^{\dot{\rho}}_{23}\, \tilde{H}_{\ad}^{\dot{\m}, \, \dot{\l} \l} + 20 \frac{x_{\bar{2} 3}{}^2}{x_{\bar{3} 2}{}^2} \bar \q^{\dot{\rho}}_{23}  {\cQ}^{\rho}\tilde{H}_{\ad}^{\dot{\m}, \, \dot{\l} \l} \non\\
&&\qquad \qquad \qquad - 20 \frac{x_{\bar{3} 2}{}^2}{x_{\bar{2} 3}{}^2} \q^{\rho}_{23}  \bar {\cQ}^{\dot{\rho}}\tilde{H}_{\ad}^{\dot{\m}, \, \dot{\l} \l} + [\cQ^{\rho}, \bar{\cQ}^{\dot{\rho}}]\tilde{H}_{\ad}^{\dot{\m}, \, \dot{\l} \l} \Big)~.
\label{SJS-z2}
\eea
From \eqref{SJS-z2}, it is now clear that only the last two terms survive after acting with $D_{(3) \a} \bar{D}_{(1) \dot{\b}}$ and then setting all the Grassmann variables to zero since neither $D_{(3) \a}$ nor $\bar{D}_{(1) \dot{\b}}$ can act on $\bar{\q}^{\dot{\rho}}_{23}$.
Then we get
\bea
&&\la V_{\b \bd}(x_1) T_{\g \d \gd \dot{\d}} (x_2) V_{\a \ad}(x_3) \ra = -\frac{1}{4}\, \frac{(x_{1 3})_{\b \dot{\m}} (x_{1 3})_{\m \dot{\b}} (x_{2 3})_{(\g \dot{\l}} (x_{2 3})_{\d) \dot{\rho}} (x_{2 3})_{\l (\gd} (x_{2 3})_{\rho \dot{\d})} }{(x_{13}{}^2)^4 (x_{23}{}^2)^6 }\non\\
&& \qquad \qquad \times  \bigg\{ 20 \ri \, \big( D_{(3) \a}\q_{23}^{\rho}\big) \cD^{\m} \bar{\cQ}^{\dot{\rho}}\tilde{H}_{\ad}^{\dot{\m}, \, \dot{\l} \l} \big| + 4 \ri \, \big( D_{(3) \a}\q_{13}^{\m}\big) [\cQ^{\rho}, \bar{\cQ}^{\dot{\rho}} ] \tilde{H}_{\ad}^{\dot{\m}, \, \dot{\l} \l} \big| \non\\
&& \qquad \qquad \qquad + \ri \,D_{(3)\a} \cD^{\m} [\cQ^{\rho}, \bar{\cQ}^{\dot{\rho}} ] \tilde{H}_{\ad}^{\dot{\m}, \, \dot{\l} \l} \big|\bigg\} + \rm c.c.
\eea
In the first two terms we will use $D_{(3) \a}\q_{23}^{\rho} =- \d^{\rho}_{\ \a}$. In the last term only the structures linear in  $\bar{\bm \Q}_3$
will contribute. We then use the identity
\be 
D_{(3) \a} \bar{\bm \Q}_{3 \dot{\a}} \big| = - X_{\a \ad}~. 
\ee

After a lengthy calculation, we obtain
\bsubeq
\label{77.44}
\bea
\la V_{\b \bd}(x_1) T_{\g \d \gd \dot{\d}} (x_2) V_{\a \ad}(x_3) \ra &=& \frac{ (x_{13})_{\b \dot{\m}} (x_{13})_{\m \bd} (x_{23})_{(\g \dot{\l}} (x_{23})_{\d) \dot{\rho}} (x_{23})_{\l (\gd} (x_{23})_{\rho \dot{\d})}}{(x_{13}^2)^4 \,(x_{23}^2)^6} \non\\
&& \times \, h_{\a \ad}{}^{\dot{\rho} \rho, \, \dot{\m} \m, \, \dot{\l} \l} (X_3)~,
\eea
where the tensor  $h_{\a \ad}{}^{\dot{\rho} \rho, \, \dot{\m} \m, \, \dot{\l} \l} (X)$ is given by
\bea
&&h_{\a \ad}{}^{\dot{\rho} \rho, \, \dot{\m} \m, \, \dot{\l} \l} (X)= 4d_3 \Big\{ \frac{2}{X^6} \d^{\l}{}_{\a} \d^{\dot{\l}}{}_{\ad} X^{\dot{\m} \rho} X^{\dot{\rho} \m} + \frac{3}{X^6} \d^{\l}{}_{\a} \d^{\dot{\l}}{}_{\ad} X^{\dot{\m} \m} X^{\dot{\rho} \rho}\non \\
&&\qquad \quad  + \frac{2}{X^6} \d^{\m}{}_{\a} \d^{\dot{\m}}{}_{\ad} X^{\dot{\l} \l} X^{\dot{\rho} \rho} - \frac{3}{X^8} X^{\dot{\m} \rho} X^{\dot{\rho} \m} X^{\dot{\l} \l} X_{\a \ad} \Big\} \non\\
&&\qquad \quad -4 (d_2 + 3d_4 +3d_3) \Big\{ \frac{1}{X^6} \d^{\l}{}_{\a} \d^{\dot{\m}}{}_{\ad} X^{\dot{\l} \rho} X^{\dot{\rho} \m} + \frac{1}{X^6} \d^{\m}{}_{\a} \d^{\dot{\l}}{}_{\ad} X^{\dot{\rho} \l} X^{\dot{\m} \rho} \Big\}~.
\eea
\esubeq
Note that due to the symmetries of the prefactor only the symmetric part in $(\l, \rho)$ and  $(\dot{\l}, \dot{\rho})$
of the function $h_{\a \ad}{}^{\dot{\rho} \rho, \, \dot{\m} \m, \, \dot{\l} \l} (X)$ contributes. 
Also note that the correlator~\eqref{77.44} is fixed up to two independent coefficients. 

Let us now rewrite  
the tensor $h_{\a \ad}{}^{\dot{\rho} \rho \, \dot{\m} \m, \, \dot{\l} \l} (X)$ in vector notation, where we define 
\bea
h^{abcd} = \frac{1}{16} (\s^a)_{(\rho (\dot{\rho}} 
(\s^b)_{\l) \dot{\l})} (\s^c)_{\m \dot{\m}} (\tilde{\s}^d)^{\ad \a} h_{\a \ad}{}^{\dot{\rho} \rho, \, \dot{\m} \m, \, \dot{\l} \l} (X)~.
\eea
The tensor $h^{abcd} $ thus introduced is symmetric and traceless in $(a, b)$. We obtain
\bea
h^{abcd} (X) &=&  \frac{d_3}{X^4} \big( \eta^{ac} \eta^{bd} + \eta^{bc} \eta^{ad} \big) + \frac{1}{X^6} (d_2 + 3d_4 -2d_3) \big(\eta^{ad} X^{b} X^{c} + \eta^{bd} X^{a} X^{c} \big)\non\\
&&-\frac{1}{X^6} (d_2 + 3d_4) \big( \eta^{ac} X^b X^d+ \eta^{bc} X^a X^d \big) + \frac{2}{X^6}(d_2 + 3d_4 + d_3) \eta^{cd}X^a X^b  \non\\
&&- 12 d_3 \frac{X^a X^b X^c X^d}{X^8} - \frac{1}{2 X^4} (d_2 + 3d_4 + 2d_3) \eta^{ab} \eta^{cd} + \frac{4d_3}{X^6} \eta^{ab} X^c X^d~. 
\label{VTV-hvect}
\eea
Our final result in the form \eqref{VTV-hvect} agrees perfectly with the general structure of the three-point function involving two vector currents and the energy-momentum tensor constructed by Osborn and Petkou~\cite{OP}. More precisely, we find the following relations:
\bea
d_3 = e~,\qquad (d_2 + 3d_4) = -c~,
\eea 
where $c$ and $e$ are the two linearly independent, real parameters appearing in the function $\tilde{t}^{abcd} (X)$ in eq. (3.13) in \cite{OP}. 


\section*{Acknowledgements}
The authors would like to thank Stefan Theisen and Arkady Tseytlin for valuable discussions. 
The work is supported in part by the Australian Research Council, project No. DP200101944. 


\begin{footnotesize}

\end{footnotesize}
\end{document}